\documentclass[a4paper,10pt]{article}
\usepackage{graphicx}
\usepackage{longtable}
\begin{document}

\begin{center}
{\bf \huge Dark Matter Macroscopic Pearls, 3.55 keV X-ray line, How big ?}\\
{\bf \bf H.B. Nielsen, Niels Bohr Institutet(giving talk),} \\
{\bf Colin D. Froggatt, Glasgow University.}\\
\end{center}

\date{Bled , July  , 2020}

\begin{abstract}
We study the 3.55 keV X-ray suspected to arise from dark matter in
our model of dark matter consisting of a bubble of a new phase of the vacuum,
the surface tension of which keeps ordinary matter under high pressure
inside the bubble. We consider two versions of the model:

\begin{itemize}
\item{Old large pearls model:}
We worked for a long time on a pearl picture
with pearl / bubbles of cm-size adjusted so that the impacts of them on
earth could be identified with events of the mysterious type that happened in
Tunguska in 1908. We fit both the very frequency, the 3.55 keV, and the
overall intensity of the X-ray line coming from the center of the Milky Way and
from galaxy clusters  with one parameter in the model in which this
radiation comes from collisions of pearls.

\item{New small pearl model:} Our latest idea is to let the pearls be smaller than atoms
but bigger than nuclei so as to manage to fit the 3.5 keV X-rays
coming from the Tycho supernova remnant in which Jeltema and Profumo observed this
line. Further we also crudely fit  the DAMA-LIBRA observation with the small pearls,
and even see a possibility for including the electron-recoil-excess seen
by the Xenon1T experiment as being due to de-excitation via electron emission
of our pearls. The important point of even our small size pearl model is that
the cross section of our ``macroscopic'' pearls is so large that the pearls
interact several times in the shielding but, due to their much larger mass than the
typical nuclei, are {\em not stopped by only a few interactions}. Nevertheless only
a minute fraction of the relatively strongly interacting pearls reach the
1400 m down to the DAMA experiment, but due to the higher cross section we
can fit the data anyway.

\end{itemize}

\end{abstract}

{\bf This article is one of the  Proceeding contributions to
the Workshop ``What comes beyond the Standard Models'' held
electronically in Bled 2020}
\section{Introduction}

The main purpose of the present article is to put forward
the latest developments of our long speculated idea that the
so far mysterious dark matter found via its gravitational
forces, instead of consisting of particles of atomic masses or an Axion-like
condensate, could consist of our proposed type of macroscopic objects
with a mass much bigger than that of genuine atoms.

We started our speculations already years ago by supposing
cm-size pearls make up the dark matter, but they will be developed
in the section \ref{small} below into the idea that these pearls
could indeed be much smaller and of geometrical size even smallish compared to
atoms, although the mass should still be appreciably larger than that
of atoms.

{\em We shall stress small macroscopic pearls.}

Even such a dramatic change in our old model into a version with
much smaller pearls would not be observed via the gravitational effects
provided just that the density of mass per unit volume is kept the same.
It is also this fact that really only the mass density matters for the
gravitational effects, that makes it possible that these effects cannot
distinguish our types of heavy or relatively lighter pearls from the
more usual assumption of only atomic weight particles, such as supersymmetric
partners of $Z^0$ or photon say in superstring theory.

However, assuming that indeed the X-ray radiation \cite{Bulbul, Boyarsky}
observed by satellites and suspected to come from dark matter does indeed
come form dark matter requires more specific models for what the dark
matter could be; e.g. it could consist of some new sort of sterile
neutrino able to decay although very seldomly into a photon and
e.g. an ordinary neutrino. Such a sterile neutrino should then
of course have a mass equal to just two times the photon energy number 3.55 keV
of the observed X-ray radiation counted in the rest frame of the supposed
dark matter in the region observed.

\begin{itemize}
\item{\bf Our Old Model:}
We develop an alternative version of our model \cite{theline, Dark1,Dark2, Tunguska}
in which dark matter consists of cm-size pearls with
masses of $10^8$ kg under the attempt to identify the X-ray radiation
seen by sattelites\cite{Bulbul, Boyarsky} and supposed to originate from dark matter with the
energy per photon 3.55 keV. We shall  discuss the possibility
that the
dark matter pearls be
much smaller but still macroscopic. This is our new model with small pearls
of a size smaller than atoms but bigger than atomic nuclei.

Actually we assume that our pearls have a skin surrounding them
keeping some ordinary matter inside the pearls under such an (appropriate)
pressure that, in the electron system of this ordinary
matter inside, there appears an energy gap between filled and empty electron
states  - called the homolumo gap (to be explained later) - of size close to the energy difference just
3.55 keV of the observed radiation. The idea then is that there can
be excitations being (loosely) bound states of an electron in one of the
lowest  empty states and a hole in one of the at first filled states.
These excitons should  have an energy close to the observed photon
energy in the line. Then one could have that the photons observed astronomically
by the satellites are photons from the decay of such excitons in the highly
compressed ordinary matter material in our model supposed to exist
inside pearls making up the dark matter.

It is a major part of our work \cite{theline}  to evaluate the
rate of such X-ray radiation that will result under the assumption that
the main production of the 3.5 keV radiation comes about when two of our
dark-matter-pearls collide with each other. We claim it to be a great success that
the magnitude of this rate of radiation can be fit together with the
energy per photon, the number 3.5 keV.

We shall in the present article have in mind really two models, which
are essentially inconsistent
with each other,
In the first model the mass of one pearl is about
$1.4 *10^8 kg$ and in the other model the mass is about
$10^4$ GeV = $10^{-23} kg$.
The old value of $1.4 *10^8 kg$ was
taken as a fit to the famous Tunguska-event in 1908 taken to be due to the impact
of one of our pearls. The small mass proposal of about $10^{-23}kg$ is rather
inspired by an attempt to fit to the DAMA (-LIBRA) experiment (by most people presumably believed to
be due to something else other than dark matter).
(A presentation of the DAMA results is given in the present Bled Workshop proceedings).

\item{\bf Observational Discussion:}

Our small mass $10^4$ GeV $\approx 10^{-23} kg$ pearl proposal
is filled with ordinary matter with an estimated density
of the order of $10^{14} kg/m^3$ as we fit the size of the pearl.
It is clear that the size of such a small pearl will nevertheless
be so big - bigger than an atomic nucleus - that the cross section
is likely to be so big that it could not possibly pass through
about 1400 m into the earth without interacting. So in this sense
our dark matter pearls are not WIMPS since the WI in this acronym
stands for {\bf w}eakly  {\bf i}nteracting. It could still be {\em dark}
in the sense
that the interaction with e.g. light {\em per mass unit} could be
small, but not small
per pearl.

With such a strong interaction one may worry whether such pearls have any
chance of reaching down to give any signal in underground experiments
looking for dark matter, because the pearls might be stopped in the
shielding above the experimental apparatus; but here the reader
should have in mind that a pearl that is heavy compared to atoms or nuclei,
when it hits, will not be stopped but just deliver a smaller part of its
kinetic energy to the hit particle, so that the latter obtains a speed
of the same order as the speed of the incoming pearl.

Of course, if one has a hugely heavy pearl as we estimated of cm-size and
with the large mass of $1.4*10^8$ kg, then it will cause a major catastrophe,
like the famous one in Tunguska, and a potential underground laboratory
would be destroyed rather than making a proper observation.

But with the small size pearl having a mass in the $10^4$ GeV range the
pearl would still interact a lot with the earth in the shielding, but
possibly not enough to be fully stopped before reaching say the DAMA-LIBRA
laboratory proper. Actually we shall imagine that a very small
fraction of the pearls come though to the laboratory, by accident so to say.

If the pearls interact several times passing through the experimental apparatus they will
be {\em disqualified} as dark matter, which is usually assumed to have so small a
cross section that they only interact {\em once} in the detector.
Even if a dark matter pearl interacts several times in the shielding
- but is not observed to interact because a high mass is not stopped
but can continue - it may well be observed essentially as a dark matter event
anyway.

Really we would like to propose a picture for the $10^4$ GeV pearl mass proposal
that a major part of the pearls end up getting stopped in the shielding
- the earth above the experimental hall underground - but that the pearls
with the smallest cross sections come through to the experimental apparatus
and are observed there. If the pearls have a much bigger cross section than
normal WIMPs they may well produce a non-negligible number of events
even if the number reaching through is much lower than the number
of WIMPs one would have expected.

In other words for the $10^4$ GeV mass pearls we shall speculate
that compared to the usual WIMP picture the much higher cross section
of our pearls than that of the WIMPs can compensate for the lower number of
pearls than of WIMPs reaching to the experimental apparatus for two
reasons:
\begin{itemize}
\item There are fewer pearls than WIMPs if the pearls are as
suggested heavier than the WIMPs, because we have to keep the gravitational
effects the same to have the same mass density in the universe.
\item There are few pearls also because some of the pearls get stopped
in the shielding due to the bigger cross section in spite of them being
heavy and not so easy to stop.
\end{itemize}

Now we should also mention that what is truly measured in the
DAMA-LIBRA experiment is not so much the full numbers
of presumed dark matter particles interacting with the apparatus, but rather
the seasonal variation of the number of events. If indeed what they see
in DAMA-LIBRA were due to our rather strongly interacting pearls, then
there would be a seasonal effect partly due to the pearls coming in one season
with higher speed than in another so they  would be able to penetrate deeper.
If by chance the
depth of the laboratory is close to the average stopping place of the pearls,
such an effect of different penetration depths in the different seasons
might be delicate to estimate, but could make it possible to get a bigger
seasonal effect than estimated in a more simple way.

Let us immediately remark, that if indeed such seasonal variation due
to relatively small changes with season of the penetration depth of interacting
pearls (dark matter particles), then this could mean that the DAMA-LIBRA
type of experiment measuring mainly the seasonal effect could be
favoured in finding a signal over other experiments not using this
technique. This would help solving the main problem or mystery
in connection with the DAMA-LIBRA experiment: Why do the other
underground experiments looking for dark matter not see the same
amount of it as DAMA-LIBRA? Now we would answer that DAMA-LIBRA
may sit close to the average penetration depth and in one season this
penetration depth is a bit deeper and DAMA-LIBRA sees a lot, while in another
season the average penetration depth is a bit higher up and one does not see so
much. Actually our fit suggests that the average penetration depth
is only a small part of the way down the 1400 m but the falling off tail
of the distribution which DAMA observes varies exponentially with the
variation in average penetration depth and a rather big seasonal effect
is indeed expected.

We want to conclude that IMPs (= interacting heavy particles) as our pearls could
be denoted rather that the usual WIMP picture is a possibility for what
the underground experiment DAMA-LIBRA could have observed.

And our argument about the penetration depth could be used to explain
that other experiments did not see the same dark matter.
\end{itemize}
\subsection{Plan}

In the following section \ref{gravitational} we present a couple of figures
about the dark matter as known already via its gravitational forces, and in
the following section we give a couple of figures about impacts of objects like meteors falling
on earth with the purpose of comparing the energy delivered with that
which dark matter could deliver, if it fell like other objects. Then in section
\ref{requisits} we review some of the ideas needed to understand our type of
model with pearls consisting of a bubble of a new type of vacuum
(this is just our speculation because so far nobody really saw any new vacuum
convincingly). In the subsections of this section we present in
\ref{MPP} our postulated new law of Nature ``Multiple Point Principle'',
which is the main new assumption in our work in as far as, except for
this multiple point principle, we only need the Standard Model as the laws of
nature. We only make further speculations on the dynamics such as the existence
of bound states or in general on results of the too hard to calculate, but
by far not excluded possibilities in the Standard Model. In the
subsection \ref{walls} we
say a few words about the domain walls that will separate such different
phases of the vacuum that we speculate exist. In subsection \ref{other} we mention
the effects other than gravitational ones which are probably due to the dark matter.
The most important such effect
for the present work is the excess X-ray radiation
observed as a tiny peak above the best
understanding fit to the X-ray spectrum at the photon energy 3.55 keV.
Other such likely dark matter effects are an excess of positrons and the associated
gamma rays; and then, what we are very keen on, one of the experiments
Xenon1T meant to look for dark matter saw a little excess of electrons
appearing in the apparatus, at first seemingly not dark matter; but we think
it could be our dark matter pearls passing slowly through and delivering
electrons with just the energy 3.55 keV.

In section \ref{usual} we mention that the type of dark matter models
most popular in the literature, except for black holes making up the dark
matter, need to modify the Standard Model by introducing extra particles
corresponding to extra fields. Most popular is to use supersymmetry
models in which there has to be included as many new physics particles
as there are particles already. Compared to that one should understand that we
only add a new fine-tuning principle the ``Multiple Point Principle'', which is
an extra assumption about the values of coupling constants that can even be checked and
at least are close to work, while the usual modified Standard Model has
lots of extra particles not yet found.

Next in section \ref{oldfit} we discuss the fitting with our
large pearl model, and in section \ref{small} we then
consider the model with the ``small'', meaning little less than atomic
size, pearls. Really this ``small'' size is in fact very large compared to what is
considered in more conventional models (such as supersymmetry).
In the subsection \ref{Tycho} we extract the ratio of cross section to
mass for the dark matter required from the observation of the 3.55 keV
X-rays from the Tycho supernova remnant and compare it to the corresponding ratio
for nuclei in subsection \ref{coincidence}. Then in the subsection
\ref{CF} we present the fit of the small pearl model.
We discuss the possibility of fitting the data from the DAMA(-LIBRA)
experiment and limits on the pearl mass in subsections \ref{DAMA} and
\ref{simple}. The Xenon1T electron recoil excess is discussed in
subsection \ref{xenon1t}.

In section \ref{conclusion} we resume and conclude the article.

\section{We know something from the gravitational studies}
\label{gravitational}
As is well-known the dark matter has mainly and in fact
possibly only been seen by its gravitational effects - and it
could still be a possibility that there is no dark matter, but instead
that something is wrong with our understanding of the gravitational
force - but even from only observing  it
via the gravitational force, one can nevertheless derive some
understanding of its distribution and velocity.

\begin{figure}

\begin{center}
\includegraphics{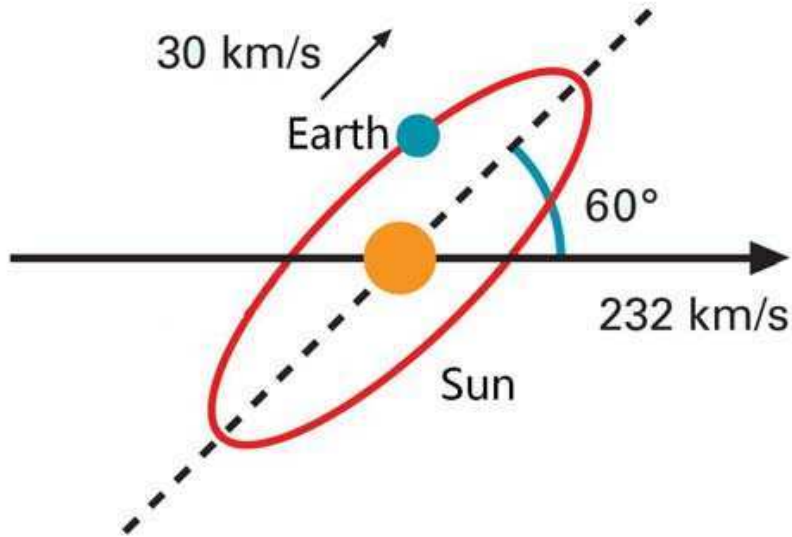}
\end{center}
\caption{{\bf Motion relative to Dark Matter:} Here is drawn
how the solar system moves along relative to the supposed rest system
of the bulk of the dark matter. One shall imagine the earth going around the
ellipse drawn which in perspective is an approximate circle representing the
orbit of the earth. Note how the speed of the earth w.r.t. the dark matter
average will vary with the season.}
\label{f1aa}
\end{figure}

In fact one can already estimate that the solar system as a whole moves
relative to the local dark matter average velocity with a speed of
232 km/s according to Figure \ref{f1aa}.

Further the distribution of the dark matter,
motion of dark matter, stars etc. provide us with:

\begin{figure}\label{f2}
\includegraphics[scale=0.8]{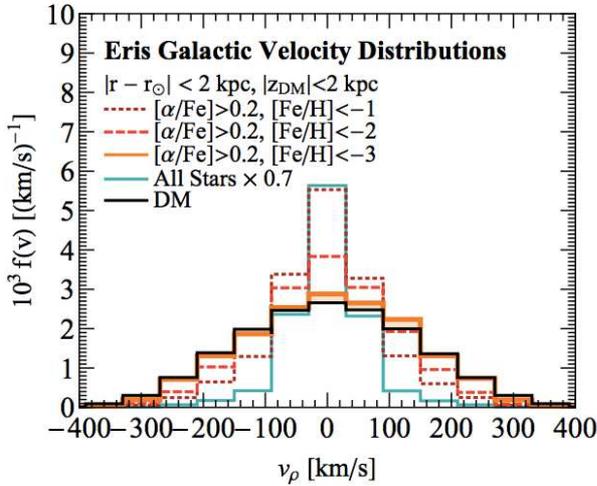}
 \caption{Velocity histogram of different components of the Milky Way, as
seen in the ERIS simulation. The black histogram shows the velocity
distribution of dark matter. The cyan histogram illustrates the velocity of
all stars, and has a much larger central peak than the dark matter
distribution. The orange histogram, however, which includes only metal-poor
stars, is very similar to the dark matter velocity distribution.
(Herzog-Arbeitman et al. \cite{Herzog-Arbeitman})}
\end{figure}

{\bf Numbers for Crude Estimates}
\begin{itemize}
\item{\bf Density of Dark Matter in Solar System Neighborhood:}
\begin{eqnarray}
D &=& \frac{0.3\, GeV}{cm^3} = 5.35 *10^{-22}\frac{kg}{m^3}
\end{eqnarray}
\item{\bf Typical Speed (also relative to each other):}
\begin{eqnarray}
v &=& 200\, km/s = 2*10^5\, m/s
\end{eqnarray}
\item{\bf Rate of Impacts on crossing Area, per $m^2$:}
\begin{eqnarray}
Rate &=& vD = 1.07 *10^{-16}\frac{kg}{m^2s}
\end{eqnarray}
\end{itemize}

These numbers may be crudely estimated by looking at the distributions in Figure
\ref{f2}, which have been gotten from the ERIS simulation
of the dark matter.
\section{Compare to Rates of Impacts on Earth}
\label{impacts}
For the dark matter we have thus found the rate

\begin{eqnarray}
Rate &=& vD = 1.07 *10^{-16}\frac{kg}{m^2s}
\end{eqnarray}
In Table \ref{tableon7} we use this $vD$ for dark matter in
our neighborhood to derive a few estimates of impact rates for
dark matter, if dark matter were indeed macroscopic particles with
the masses listed in the first column of this table:

{\bf Hitting Rates for some Masses:}

In the first column is given the mass of the dark matter pearl.
The second column gives the rate of impacts such a mass would give per $m^2$
and in the third column this rate is translated into the time between
the impacts on this square meter. The fourth and fifth column similarly
give the rates and the time in between impacts for impacts on the Earth in total
instead of just on a square meter. Notice that in the row
corresponding to the mass of the dark matter particle being $10^8 kg$
there is - in the last column -
about 100 years between the impacts. Now it was approximately 100 years ago when
the famous Tunguska event occurred, meaning that if the Tunguska
event should be caused by a dark matter pearl, then the mass would be of the
order of $10^8kg$.

\begin{table}[h]
\begin{tabular}{|c|c|c|c|c|}
\hline
mass & $m^2$ & $m^2$ & earth &earth\\
& rate & time & rate & time\\
\hline
$10^{-16}kg$& $1 s^{-1}$ & 1 s& $5*10^{16}s^{-1}$& $2*10^{-15}s$\\
$=5*10^{10}GeV$&&&&\\
\hline
$10^{-8}kg = 10 \mu g $&  $10^{-8}s^{-1}$&$10^8 s=3y.$& $5*10^8s^{-1}$&
$2*10^{-9}s$\\
\hline
1 kg & $10^{-16}s^{-1}$ & $10^{16}s$& $5s^{-1}$&$0.2 s$ \\
\hline
$10^8 kg= 10^5 ton$& $10^{-24}s^{-1}$& $10^{24}s$&$5*10^{-10}s^{-1}$& $2*10^9s$\\
&&&&$\sim 100 y$\\
\hline
\end{tabular}
\caption{ A few rates for hypothetical dark matter pearls}
\label{tableon7}
\end{table}

Next we now give a similar table for meteor impacts as observed, impacts a
priori expected to be made from ``ordinary matter''( i.e. atoms). Here it is
meant that the impacts are counted for the whole Earth:


{\bf Compare Impacts of Ordinary Matter}

$10^{-2}\, kg$ : $10^5 $ per year\\

1 kg : $10^4 $ per year.\\

$10^8\, kg$ : $10^{-3}$ per year.\\

You may consider the numbers in this table \ref{tableon7} as extracted from
Figure \ref{f3}.

Since a year has $3.16*10^7$ s this corresponds
to a mass density $ D_{meteors}$ times the velocity $v_{meteors}$ being
of the order
\begin{eqnarray}
v_{meteor}D_{meteor} &\sim& \frac{10^4\, kg/year/eartharea}{3.16*10^7 s/year}\\
&=& \frac{3*10^{-3}\, kg/eartharea/s}{0.5*10^{15}m^2/eartharea}\\
&=& 2*10^{-18}\, kg s^{-1}m^{-2};
\end{eqnarray}
 formally a {\em factor 50 smaller} than the dark matter.
Rather than the mass of the impact object you might use its
size and then we get the graph in Figure \ref{f3}:
\begin{figure}
\includegraphics{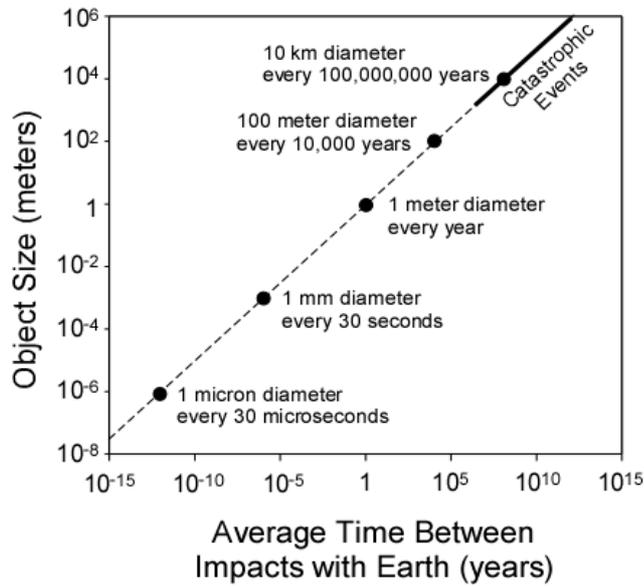}


\caption{{\bf Size of Impact goes as square root of ``time in between''}}
\label{f3}
\end{figure}

From this figure \ref{f3} we can read off an approximate dependence
of the size of the impacts on earth and their frequency. Approximately the
inverse frequency being the ``time between'' goes as the square
of the size of the impacting object. So a formula easy to remember is:
\begin{eqnarray}
\hbox{``impact size'' in m} &=& \sqrt{\hbox{av. ``time between'' in years}}
\end{eqnarray}
on earth.

\begin{figure}
\includegraphics{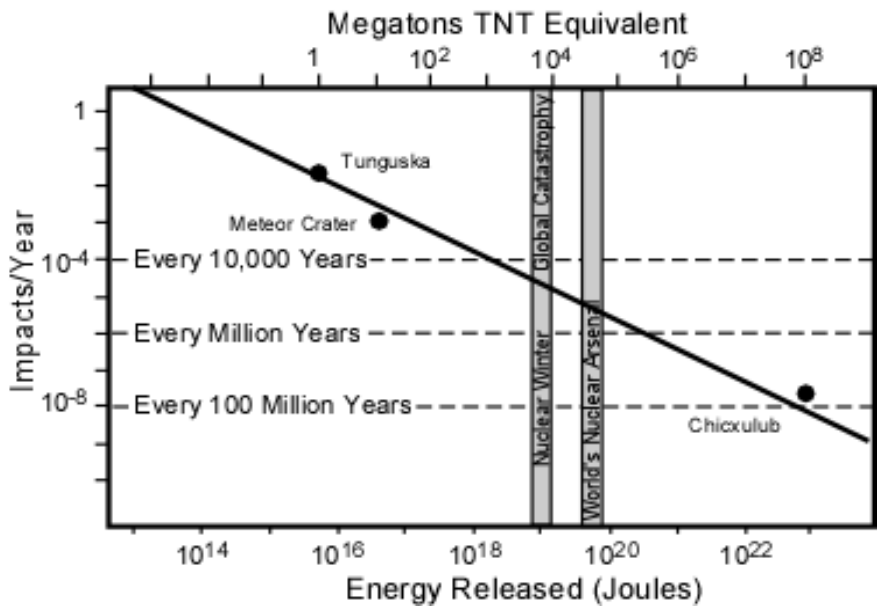}
\caption{Relation between energy released and impacts per year}
\label{f4}
\end{figure}

On Figure \ref{f4} we see the relation between energy release by the
impact and again the frequency measured in impacts per year.

{\bf Would Macroscopic Dark Matter Dominate Meteors?}
\begin{itemize}
\item Taking very roughly the graph as having the slope
-1 in the logarithmic plot we may read off that the energy of
impacts per year is of the order of magnitude of $10^{13} J/y $ to
$10^{14}J/y$.
\item To compare with that the kinetic energy in the column of dark matter
hitting the earth per year is for non-relativistic dark matter particles of
the order of
\begin{eqnarray}
&&\hbox{``dark matter power on earth''}=\\
&=&\frac{1}{2}* (300\, km/s)^3*0.3GeV/cm^3 *
\pi *(6.38 * 10^6m)^2\\
&=& \frac{1}{2} (3*10^5m/s)^30.3*1.78 *10^{-21}kg/m^3*\pi*(6.38*10^6 m)^2\\
&=& 1.27*10^{16} J/y
\end{eqnarray}
(using 1 year = 31556952 s)
\end{itemize}

So it looks that unless some of the kinetic energy of the dark matter
hitting the earth is lost from showing up as observable impacts, there is
too much energy in the dark matter to match the impacts as observed.

In our old work \cite{Tunguska} we took it that because of the
smallness of even cm-sized pearls they penetrate so deeply into
the earth that it is realistic that an appreciable part possibly 19/20
of the energy is penetrating so deep into the earth, that it does not
appear as observed energy on the surface of the earth. Since we could well
find it consistent that our big pearl (=cm-size) would go thousands of km
into the earth, it would indeed be hard to get all the energy out
so quickly as to be identified with the energy of the impact.

\section{Requisites for Our Model(s)}
\label{requisits}
Before going on to fit our type of model and discussing how
well such pearl models for the dark matter matches much of our knowledge
about the dark matter, as it actually will, we shall put forward
a few prerequisites needed for understanding the speculations making up
at least one concrete example of a macroscopic pearl model of the
dark matter.

As a motivation for just our concrete picture for how the pearls could come
about let us stress: {\em Our picture of dark matter pearls can come
about in the pure Standard Model, i.e. without any new physics in the sense
of new basic particles.} We shall rather only speculate about new particles
which are bound states of the already known particles, and thus do not
require any modification of the Standard Model. We have e.g. no
supersymmetric partners, because we do not have supersymmetry at least
not in the relevant region of energy for our model.

Gia Dvali showed that the existence of several vacua is inconsistent unless they
are degenerate in the article ``Safety of Minkowski Vacuum''
\cite{Dvali}.

\subsection{Multiple Point (Criticality) Principle}
\label{MPP}
The point in our work, which comes closest to assuming new physics,
is the principle that the coupling constants of the true model for physics
- for our purpose here the Standard Model - are by a ``new Law of Nature''
tuned in to just arrange that there are a series different phases of the vacuum
- different vacua we could say - which all have the same energy density
( = cosmological constant) \cite{MPP1,MPP2,MPP3,MPP4}. We call this principle
of such
fine-tuning of the coupling constants the
{\em Multiple Point (Criticality) Principle} (MPP) \cite{MPP1,MPP2,MPP3,MPP4}. There
have been given various
arguments for it \cite{Dvali,MPP1,MPP2,MPP3,MPP4}, and we can claim that
using it we have even made correct
{\em predictions}, e.g. the number of families,
prior to the LEP measurement of the number of light neutrino species. We
fitted fine structure constants in a rather complicated model
called ANTIGUT and the fitting parameter was indeed the number
of families. We {\em pre}dicted that to be 3. Later  we obtained a
mass prediction \cite{tophiggs} for the
Higgs of $m_{Higgs}=135 \pm 10$ GeV {\em before the Higgs was found}.

For our pearl-models of the dark matter it is important that Nature should
have this fine-tuning at least to an appreciable accuracy making the
inside and the outside vacua for our pearls of equal energy density.
This is because
otherwise almost certainly one of the phases would spread out and it would
be very hard to get pearls that are stable. Actually even with the
degenerate vacua we have in our model the need for getting the pearls
filled by ordinary matter under high pressure to withstand the pressure coming
from the tension of the surrounding skin or domain wall. Guesses as to the
order of magnitude for what the energy density difference should be, if not
tuned to be small, would be so high that our model would become unlikely.
Though, if e.g. the energy density difference was only of the order
corresponding to the observed order of magnitude for the vacuum energy
in the universe it would contribute so little over one of our pearls that
it would not disturb our calculations taking the difference to be zero.

\subsection{Domain walls in general}
\label{walls}
There is also a discussion of walls in another article \cite{Murygin} in these
proceedings.

We ourselves like to point out, that once we have the ``Multiple Point
Principle''
we have in principle the possibility that some even large
regions in space could be filled by one phase while another region
could be filled by another phase of the vacuum. Had we had
a spontaneously broken discrete symmetry it would induce a case
of ``Multiple point principle'' in as far as two or more phases related by the
broken symmetry would of course have for symmetry reasons the same energy density.
It is however, not such a case of a spontaneously broken discrete symmetry,
which {\em we} imagine in our model. We rather speculate that two a priori
different, and not connected to each other by symmetry, vacuum phases are to
be used. Having the spontaneously broken discrete symmetry is also
phenomenologically badly working, in as far it would typically lead to
random vacua coming to dominate in various regions outsides the horizons
of each other. Such outside each others horizon different dominating
vacua would cause domain walls extending over longer distances than
the horizon and in turn make up huge amounts of domain
 walls in cosmology.
Unless the wall tension was extremely small such horizon scale walls would
get to dominate under all circumstances in the long run; and that would
spoil our cosmological models.

So we must hope, and we actually do expect, that the domain walls
due to the asymmetry between their sides - i.e. due to the fact that
the different vacuum phases are not connected by symmetry -
will contract a bit more towards diminishing one vacuum than the other one.
Thus at an early stage in the history of the universe  one of the vacua
only  survives in {\em small} bubbles compared to the universe size.
It is such small surviving bubbles that should be the dark matter.
Actually even the small bubbles only survive because at a stage they
get stopped from contracting by having collected so many nucleons inside
that they can provide a sufficient pressure to stop the contraction.

For our cm-size pearls we had an estimate that the contraction of the pearls
to the stability point where they just have the size given by their content
of nucleons, counteracting the pressure, would end about the time
in cosmology, when the big bang nuclear synthesis is about to start and
the temperature is of MeV size. It is very needed for our model that
the pearls have become so compact and effectively disconnected
from the rest of the plasma before the big bang nuclear synthesis properly
begins, because otherwise our model would modify this big bang nuclear
synthesis,
and it would be an unconvincing refitting even if we managed to fit the
abundances of the various light isotopes resulting form the big bang
nuclear synthesis.

Nevertheless one should of course investigate astronomically if some
of the big voids observed in the matter distribution should actually be
a result of domain walls. If one had, for some accidental or other
reason, an astronomical size region with the same vacuum as
inside our pearls, formally an enormously large dark matter pearl,
then we would expect there to be the same matter density inside this
huge pearl as on the average in the universe. But now there would be
no way to have true dark matter in the region, because the whole
region is already formally dark matter. Pearls inside it of the present phase
vacuum would repel rather than attract nucleons and would thus totally
collapse. Therefore in such regions one would in practice lack the dark matter
and have it replaced by a higher density of ordinary matter. The latter
would, however, have electrons staying relativistic longer than
dark matter would have stayed relativistic. Thus these regions
would presumably develop their inhomogeneities later than the
regions where the present vacuum dominates. This could then be likely to delay
the development of stars and galaxies in such formal huge dark matter
bubbles of astronomical size. Such regions  might appear as voids?

\subsection{Non-gravitational Dark Matter Observations}
\label{other}
We believe it is true to say that all non-gravitational signs from
dark matter are somewhat doubtful. Nevertheless our main
aim in this article is to look especially for whether our model
can get support from the observations of one of the presumed non-gravitational
observations of dark matter, the 3.55 keV X-ray radiation in
outer space, mainly seen\cite{Bulbul, Boyarsky} from our Milky Way Center or from big clusters
of galaxies.

\subsubsection{The 3.55 keV X-rays}
\label{355}
We have already mentioned this for us so important X-ray observation
in a line of frequency 3.55 keV, which seems not to be explained by the
atomic ion transitions expected in the plasmas from which the X-rays
come. But it is only a tiny little deviation from the main fit of the X-ray
spectrum and e.g. an unexpectedly high abundance of potassium in the plasmas
could make a line in the region of the 3.55 keV be increased so much as
to replace the tiny suspected dark matter line.

Using the expectations from the gravitational knowledge about the
distribution of the dark matter, fits have been made to the
3.55 keV radiation expected both under the assumption that the emission
from a region depends linearly on the density $D$ of dark matter
and under the assumption, that the amount of 3.55 keV line radiation is proportional
to the square of the dark matter density $D^2$. It is the latter dependence
that should come out of our model, because we postulate that the 3.55 keV
radiation arises when our pearls {\em collide}. Both types of fit are not
hopeless, and even the rather well fitting analysis by Cline
and Frey \cite{FreyCline}, which
we use in our work, has at least one severe discrepancy: one of the
measurements in the outskirts of the Perseus Cluster delivers
about 1000 times more 3.55 keV radiation experimentally than one should
expect by extrapolating the fits to the other observations.

In our use of the analysis of Cline and Frey, we simply had to
delete this observation to obtain a meaningful average for the overall
scale of the radiation which is then what we ourselves sought to fit.

We should investigate, if we could understand this deviating measurement
in the Perseus Cluster as due to our pearls getting energy for 3.55 keV
radiation in a different way than from the collisions. In fact we have similar
problem with the Tycho supernova remnant in which the square of the density
$D^2$ over the supernova remnant region is very tiny in comparison
to galaxy clusters and the Milky Way Center extensive volumes.
The supernova remnant region, even taking into account the
closeness of the Tycho supernova remnant, is so small that it would
not be expected that Jeltema and Profumo should have seen the
3.55 X-ray line from the  dark matter there. But in fact Jeltema and Profumo
\cite{Jeltema}
{\em have seen 3.55 keV radiation from the supernova remnant.}

Our suggestion is that the cosmic rays or X-rays in the Tycho supernova region
can excite the pearls, which then whatever the excitation energy
- collision or cosmic ray  excitation - will emit an appreciable
part of the energy as 3.55 keV radiation.

One could of course hope - and we hope to find out - that
there are some similar cosmic rays or X-rays reaching the outskirts of the
Perseus Galaxy Cluster.

Of course, if the cosmic ray or X-ray activity is about the same in two
neighboring regions in say the Perseus Cluster, then the ratio of the
X-ray or cosmic ray feeded radiation relative to the one feeded by the
collisions will go in the ratio $\frac{D}{D^2 } =D^{-1}$. This is because the
rate from cosmic ray feeding goes as
$D *\hbox{``density of cosmic rays''}$, while the collision rate goes as
$D^2$. In the outskirts of the cluster the density of dark matter
$D$ presumably goes down, and thus the  cosmic ray feeded radiation
becomes relatively more important.

\subsubsection{Positrons and Other Gamma-rays}
\label{positrons}
Also positrons above some 10 GeV in energy have shown an excess
suggested to be due to dark matter together, as one could imagine, with
gamma-rays not in a line but in a broader spectrum. In this connection there
is a little problem:

Using usual types of model for dark matter identified with some type of particle
simply decaying into among other things the positron to make the excess,
it is very hard to avoid that associated with this positron emission one does
not also get some gamma-rays. Now, however, the fitting does not
go well and it seems that experimentally there are not so many
gamma-rays as is almost unavoidably needed for matching the positron excess!

This little tension with an elementary particle dark matter interpretation
could provide support for our type of model, because at the collision and strong
heating up of the uniting pearls a large amount of electrons will be emitted
and can
easily create electric fields that in a rather low acceleration way
can accelerate e.g. positrons. Thus one can get positrons which are not
produced at high speed almost abruptly, but which are ``slowly'' accelerated.
The latter gives much less electromagnetic radiation and thus our model
has the potential of making positrons with much fewer gamma rays connected
with them. This would agree better with the too few
observed gamma-rays.

\subsubsection{Xenon1T Electron Recoil Excess}
\label{Xenon}
Yet another effect, which we shall count as a non-gravitational effect of
dark matter, but which is not obviously dark matter at all: the
Xenon1T electron recoil excess.

Apart from the DAMA/LIBRA and the DAMA experiment all other experiments
seem to find only negative results, when looking for the dark matter
in direct searches. There was, however, found one unexpected result
\cite{Xenon1Texcess} although
at first not seemingly related to dark matter:

The experiment Xenon1T investigated what they call electron recoil
in their Xenon experiment. In the Xenon experiment one has a big tank
of liquid xenon with some gaseous xenon above it and photomultipliers
looking for the scintillation of this xenon, the philosophy being
that a dark matter WIMP e.g. hits a nucleus inside the xenon
and the recoil of this creates a scintillation signal S1 and also an electron,
which is then driven up the xenon tank by an electric field and at the end
by a further electric field made to give a signal at the top S2. By the
relative size of the signals S1 and S2 one may classify the events
- which are taken to be almost coinciding pairs of these signals S1 and S2 -
as being nucleus recoil or electron recoil. One expects to find
the dark matter in the nucleus recoils, since a dark matter particle
is not expected to make an electron have sufficient energy to make an
observable electron recoil event.

But now by carefully estimating the expected background, the Xenon1T
experimenters found an {\em excess of electron recoil events}.

Ideas proposed for explaining it include axions from the sun or neutrinos having
bigger magnetic moments or perhaps less interestingly that there could be more
tritium than expected in the xenon.

But here with our model of relatively stronger interacting particles
able to radiate the line 3.55 keV when excited we have a possible explanation:

Going through the earth above the detector and the rest of the shielding, the pearls or
particles  get excited so as to emit 3.55 keV X-ray just as they
would do in the Tycho supernova remnant, where they also get
excited by matter or cosmic rays. But then the particles passing through the
deep underground Xenon1T experiment are already excited and
prepared for emitting the 3.55 keV radiation. Now they could possibly
simply do that in the xenon tank or they might dispose of the energy
by a sort of Auger effect by rather sending out an electron with an
extra energy of 3.55 keV. Such an electron with an energy of a few keV
could be detected and taken for an electron recoil event in the
Xenon1T experiment.

It is remarkable that the signal of these excess electron recoil events
appears as having just an energy of the recoiling electron very close to the
value 3.55 keV. Indeed the most important bins for the
excess are the bins between 2 and 3 keV and the bin between 3 and 4 keV.

So we would claim that there is in our model no need for extra
solar axions or a neutrino magnetic moment, nor tritium.

But we claim it to be 3.55 keV radiating dark matter one sees in the xenon
experiment!

\subsubsection{The Dark Ages, 21 cm line}
As a possible place to look for information about dark matter
- especially of the pearl type say - is the influence it could have had
in the ``Dark ages'' before the stars lit up the universe, a
time that may be investigated through the study of the H1 radio line
of 21 cm wavelength. Recent studies \cite{DarkAge1, DarkAge2} were pointed out to us
by Astri Kleppe.

\subsubsection{Supernova Introductional Burst}
As an interesting possibility for
studying our dark matter pearls astronomically, we should also mention
our older work, 
in which we claim \cite{supernova} that our dark matter
pearls can not only help the supernovae to explode more, which is
what is called for, but also to explain a neutrino burst appearing
some hours before the genuine explosion, as appears to have been observed
by the neutrino experiment LSD \cite{MontBlanc}.

\section{Status of Searches}

Before going on to describe our models for dark matter being pearls of
a new phase of vacuum, let us shortly review the status of the searches for
dark matter in underground experiments. The plot in Figure \ref{searches} shows the excluded
regions in the cross section versus mass plane for
dark matter particles in the usual WIMP-theory:
\begin{figure}
\includegraphics[scale =0.7]{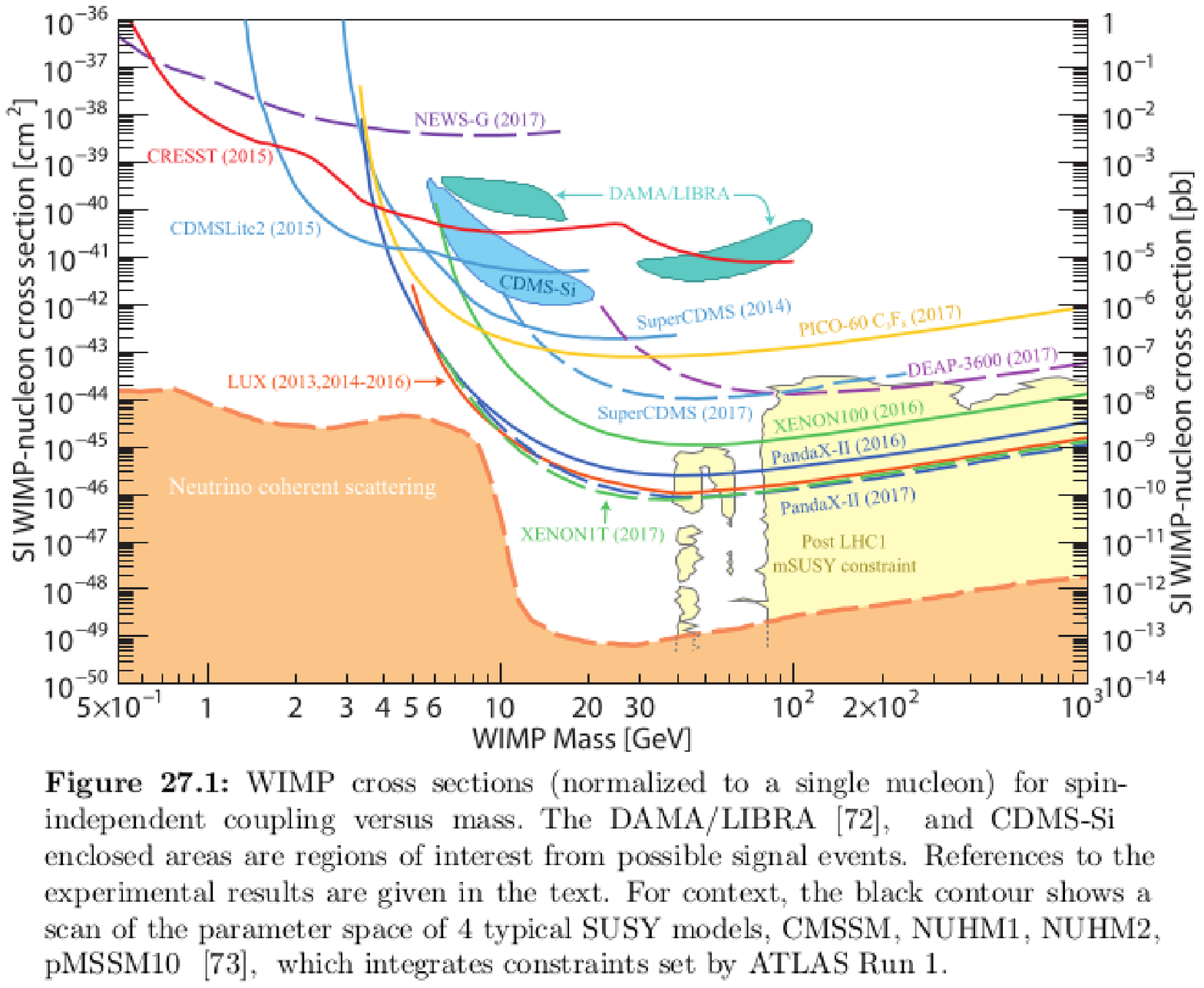}
\caption{Areas of the cross section versus mass of WIMP dark matter
particles above the curves are excluded.  So one
sees that regions favoured by DAMA and CDMS-Si are seemingly in disagreement
(although not in a theory independent way). See reference \cite{pdg2018}.}
\label{searches}
 \end{figure}
It is important to notice for our work below that inside the region excluded by
several experiments there is a spot in which the DAMA-LIBRA experiment
- in fact by 9 standard deviations - claim to have {\bf found} the dark matter
(or at least something with very similar properties) by their special
technology of looking for seasonal variations, that should appear because
the speed of the Earth relative to the average
velocity of the dark matter varies with season (see Figure \ref{f1aa} above).

\section{ Dark Matter with only the Standard Model (except MPP)}
\label{usual}
Contrary to everybody else, except for the people who take
primordial black holes for dark matter, we want to propose a
dark matter model {\em inside the Standard Model}, only with a certain
assumption about the coupling constants in the Standard Model,
that there are several vacua fine-tuned to have the same energy density. So
we have very little ``new physics'':
\begin{itemize}
\item We assume a law of nature - of a somewhat unusual kind -
the ``Multiple Point Principle'' saying: there are {\em several} different {\em
vacuum phases}, and they all have the {\em same energy density}
(or we can include
that they have $\sim 0$ energy density.)
\item Apart then from mentioning an attempt mainly with Yasutaka
Takanishi to explain the baryon excess, we shall use {\em only
the Standard Model}, even for dark matter!
\end{itemize}


\section{Our Fit}
\label{oldfit}
We performed a detailed fit with the model \cite{theline}
in which we first of all looked for the absolute scale of the intensity
of our model of dark matter pearls or balls emitting the X-ray line with photon
energies of 3.55 keV in the rest system as apparently observed by satellites
etc.

\subsection{The Intensity}
The intensity we take in our model to be emitted by pearls, that have collided
with one another - a rather infrequent event - but when they finally
collide it is assumed, that the very strong skin surrounding the
pearls can contract and thereby deliver energy, which can be used for the
radiation in the 3.5 keV line or for other frequencies. There is in our model
so to speak an active ``energy production from the contraction''. But this we
can in fact estimate, if we have the parameters of the model. Of course the
fact that we need collisions of a pair of pearls to get the radiation
in the 3.5 keV line means, that the intensity resulting in a given
region of the space becomes proportional to the {\em square } of the density
$\rho_D$ of dark matter in that region. A fit to a model of this
kind - which would also be applicable for a model in which
the dark matter particles annihilate with each other - was performed
using the astronomical - mainly satellite - data by Cline and Frey
\cite{FreyCline}. For the purpose of our model we can interpret it that they
measure an intensity proportional parameter, which basically is in our language
$\frac{N\sigma}{M^2}$, where $M$ is the mass of the typical / average
pearl, $\sigma$ the cross section for one such pearl hitting another one,
and $N$ the number of 3.5 keV photons emitted when such a collision actually
happens. From the results of Cline and Frey we find the number
\begin{eqnarray}
\label{average}
\left ( \frac{N\sigma}{M^2}\right )_{exp} &=& (1.0 \pm 0.2)*10^{23}\, cm^2/kg^2\\
&=& (8.1 \pm 1.6)*10^{-4}\, GeV^{-4}= (1.7*10^{-1}\, GeV^{-1})^4 \label{intensity-exp}\\
&=& \frac{1}{(5.9\, GeV)^4}.
\end{eqnarray}
(Here we used 1 kg = 5.62*10$^{26}$ GeV and 1 cm = 5.06*10$^{13}$ GeV$^{-1}$,
so that $\frac{cm}{kg}$ =9.00*10$^{-14}$ GeV$^{-2} $ and thus
$\frac{cm^2}{kg^2}$ = 8.10*10$^{-27}$ GeV$^{-4} $.)

Or rather we extract this number from their table:


\begin{table}[h!]
\caption{This table is based on the table 1 in reference \cite{FreyCline}.}
\begin{tabular}{|c|c|c|c|c|c|}
\hline
Name&$N<\sigma_{CF} v>*$&$v$&boost&$(\frac{N<\sigma_{CF} v>}{v *boost})*$&Remark\\
&$\left ( \frac{10GeV}{M} \right )^2$&&&$\left ( \frac{10GeV}{M} \right )^2$
&\\
Units&$10^{-22} cm^3s^{-1}$&$km/s$&&$10^{-27}cm^2$&\\
\hline
Clusters\cite{Bulbul}&480 $\pm$ 250 & 975& 30& 0.016 $\pm$ 0.008&\\
Perseus\cite{Bulbul}& 1400 - 3400&1280& 30& 0.037 - 0.09&\\
Perseus\cite{Boyarsky}& (1 - 2) $*10^5$&1280&30&2.7 - 5.3&ignored\\
Perseus\cite{Urban}&2600 - 4100& 1280& 30& 0.07 - 0.11&\\
CCO\cite{Bulbul}&1200 - 2000& 926& 30&0.04 - 0.07&\\
M31\cite{Boyarsky}&10 - 30(NFW)&116&10& 0.0086 - 0.026 &\\
&30 -50 (Burkert)&&&0.026 -0.043&\\
MW\cite{Boyarsky2}&0.1 -0.7 (NFW)&118&5& 0.00017 - 0,0012&ignored\\
&50 -550 (Burkert)&&&0.084 - 0.93&in average\label{table0}\\
\hline
Average&&&&0.032$\pm$ 0.006&\\
\hline
\end{tabular}
\end{table}

It should be noticed though that something is not fitting well in the
case of the Perseus Cluster in as far as one measurement in the outskirts of
this galaxy cluster turns out to give a factor 1000 more radiation in the
3.5 keV line than the one that would have fitted with the proportionality
to the squared density estimated from gravitational considerations.
In our averaging we left this observation out totally, since it would
have
led to a very bad fitting for the other observations. But without this
badly fitting observation we get the average (\ref{average}).

\subsection{The Frequency}

The very frequency or the photon energy 3.5 keV,
we sought to fit with
the ``homolumo gap'' in the ordinary material under high pressure
- comparable to that in white dwarf stars - inside our dark matter pearls.
Such a ``homolumo gap'' is a very general feature for materials containing
a degenerate Fermi sea of
fermions, say electrons, and in addition has some structure
- like a glass or almost all materials - consisting in that the
material in detail adjusts so as to partly lower the energy density of the
Fermi-sea. It is obvious that the energy of the Fermi sea is lower the
lower in energy the filled fermion states, whereas lowering the energy of the empty
states does not lower the total energy. The adjustment to a ground state of the
material will therefore (almost) unavoidably lead to a lowering of the
filled states and thus cause a gap between the filled and the empty states.
It is this gap between the filled and the empty single particle states which is called
the homolumo-gap. It is namely the gap between {\bf h}ighest {\bf o}ccupied
{\bf m}olecular {\bf o}rbit (the chemist expression for single
particle fermion state), HOMO and the {\bf l}owest {\bf u}noccupied
{\bf m}olecular {\bf o}rbit, LUMO.

We estimated in \cite{theline} the value in energy of
this homolumo gap partly just by a dimensional argument and partly
by using a Thomas-Fermi approximation.

The formula for our estimate of the homolumo gap, which also turns out
to be the expected frequency or photon energy for the line, was
\begin{eqnarray}
E_H
& =&
\sqrt{2}\left(\frac{\alpha}{c}\right)^{3/2} E_f.\label{EHW}
\end{eqnarray}
 Here $\alpha$ is the fine structure constant considered for the
purpose of our dimensional arguments as a velocity (by multiplying it by
the velocity of light $c$) and $E_f$ is the Fermi energy of the electrons
in the hard compressed material inside our pearls.

\subsection{The fitting and theoretical speculations}

In our model we imagine  that there are at least two phases of the vacuum
- in addition
presumably to several other ones too, but in the work now being reviewed
we cared for only two important ones - and that the one in which we do not live,
but which is realized inside the dark matter pearls, is distinguished
from the present vacuum by there being a (boson) condensate of a
speculated bound state of 6 top plus 6 anti-top quarks.
In the vacuum phase inside the pearls
we would at first have speculated that the expectation
value of the Higgs field should go to zero, but that would give us an
estimate of the tension of the skin separating the interior and the exterior
of the pearls, which would not give an acceptable fit. Indeed assuming that the usual Higgs
spontaneous breakdown of the weak gauge symmetry in the vacuum
inside the pearl is absent would suggest an order of magnitude of
the tension in the skin of the pearls of the order of
$(100\, \hbox{GeV})^3$, but the fitting we made gives an appreciably smaller tension.
\begin{table}[h]
\begin{tabular}{|c|c|c|c|c|}
\hline
Name & $\frac{\xi* 10 MeV}{\Delta V}$ &$\ln{\frac{\xi* 10 MeV}{\Delta V}}$&Uncertainty&\\
\hline
Frequency ``3.5keV''& 5.0&1.61 &100\%&\\
Intensity $\frac{N\sigma}{M^2}$& 3.8 & 1.3 &90\%&\\
 $S^{1/3}$ theory 1)&0.28&-1.3&40\%&\\
 $S^{1/3}$ theory 2)&1&0&40\%&\\
Combined theory $\xi$, $\Delta V$&0.18&-1.7&100\%&\\
\hline
Ratio $\frac{t_{spread}}{t_{radiation}}$=1 & 2.4& 0.88& 80\%&l.b.\\
\hline
 \end{tabular}
\caption{Table of four theoretical predictions of the parameter
${\frac{\xi* 10 MeV}{\Delta V}}$ on which the quantities
happen to mainly depend.
The first column denotes the quantities for which we can provide a theoretical
or experimental value to be expected for our fit to that quantity.
The next column gives what these expected values need the
parameter combination ${\frac{\xi* 10 MeV}{\Delta V}}$ to be.
The third column is the natural logarithm of that required value for
the ratio ${\frac{\xi* 10 MeV}{\Delta V}}$, i.e.
$\ln{\frac{\xi* 10 MeV}{\Delta V}}$. The fourth column
contains crudely estimated uncertainties of the parameter
thus fitted counted in this natural logarithm.
In the last column we just marked the ratio $\frac{t_{spread}}{t_{radiation}}$
with l.b. to stress that it is only a lower bound and shall not be considered a
great agreement for our theory. }
\label{table2}
\end{table}

\begin{figure}
\begin{center}
\includegraphics[scale=0.4,angle=270]{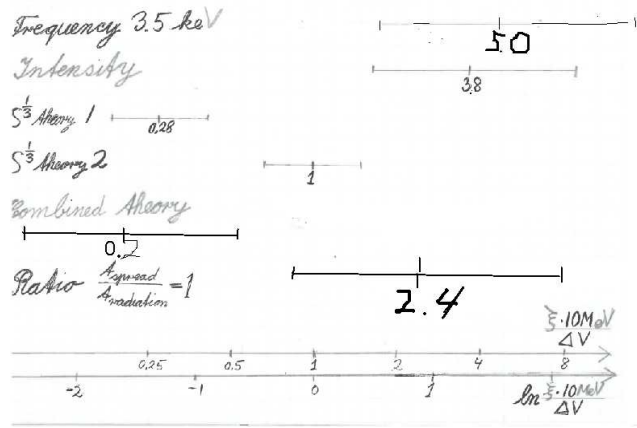}
\end{center}
\caption{The values of the ratio ${\frac{\xi* 10\ MeV}{\Delta V}}$
as needed for four constraints.
There are two experimental constraints from the frequency and intensity of the 3.5 keV radiation respectively
and two theoretical constraints in two versions corresponding to taking theory 1 or theory 2
for the tension.
We make the simplifying assumption that all energy from the surface contraction in a collision
gets emitted as 3.5 keV X-rays.
The sixth line ``Ratio $\frac{t_{spread}}{t_{radiation}}=1$'' represents the
lower bound ensuring that
all the energy actually
goes into 3.5 keV radiation.
}
 \label{fig:figure1}
\end{figure}

The essential parameter we used in our fit was defined as
\begin{eqnarray}
\frac{\xi *10 MeV}{\Delta V}&=& \frac{10 MeV * R/R_{crit}}{
\hbox{``potential difference for nucleon in the two vacua''}}.
\end{eqnarray}
In order to reduce the number of parameters in our earlier paper \cite{Tunguska}
we assumed that the pearls just had such a size that they were on the borderline to
collapse and we call the radius of such barely stable pearls
$R_{crit}$. We now denote the actual radius of the (typical) pearl by
just $R$ and define the parameter $\xi = \frac{R}{R_{crit}}$.
The parameter $\Delta V$ is the binding energy of a nucleon
relative to when it is in the vacuum phase in the interior of the pearls.
One should
imagine that nucleons are attracted by the pearls by having a lower
potential by the amount $\Delta V$ inside the pearl. If the
pearl gets too small and the pressure from the skin thus too high
it will pay energetically for the nucleons inside the pearl to escape
and the pearl thus collapses; this is what
happens when the radius is smaller than the critical radius $R_{crit}$.
The $10\, \hbox{MeV}$ was just a conventional number, we put in to make
the parameter dimensionless.

It turned out from our calculations that the combined parameter ratio
$\frac{\xi*10MeV}{\Delta V}$ is the main one to fit, because the interesting
measurable and theoretically interesting quantities mainly depend on it.

We thus used it to make fits especially to the experimentally predictable quantities,
the intensity of the 3.5 keV radiation scale and the very frequency
3.5 keV.
The fitted values of the
combined parameter $\frac{\xi *10MeV}{\Delta V}$ for these quantities
are presented in Table \ref{table2} together with those
expected for a tension of $(100\, \hbox{GeV})^3$ as obtained from
the Higgs field consideration above (theory 1)
- even a somewhat smaller value for the tension is speculated about and called
theory 2 -
and for a theoretical expectation. These predictions are also plotted in
Figure \ref{fig:figure1}.

We see that the theoretical expectations for the tension $S$ tend to
fit with too small values of our parameter combination
$\frac{\xi*10MeV}{\Delta V}$ and so does our theoretical estimate
of the $\xi$ deviation from criticality combined with the expected value
for $\Delta V$ represented in the table and the figure below as ``combined
theory''. The last line in the table and the figure represents
a parameter value below which it is expected that more and more
energy is lost to higher frequency radiation than the 3.5 keV one. This is because
the pearl in the collision gets heated up and then the heat spreads out so quickly
that only a very little part goes into the line observed as the
3.5 keV line. The point is indeed that we expect the temperature from
the contraction of the surface to be much higher than 3.5 keV, but then this
heat spreads out of course gradually on a second time scale to the whole pearl.
Under this spreading out there is a spreading border at the place to which
the heating has reached at any moment. Near that border the temperature
is about 3.5 keV and the 3.5 keV radiation is produced and because
the pearl material is supposed to be transparent to the 3.5 keV and
lower frequency radiation, it is radiated out to outer space. But if  the
heat reaches all through the pearl the outer surface of the pearl gets
appreciably hotter than 3.5 keV; then most radiation comes with higher
frequency and is correspondingly lost for radiation in the observed
3.5 keV line. The ``time ratio $\frac{t_{spread}}{t_{radiation}}$ =1''
represents the fitting to the value 2.4 of our parameter
$\frac{\xi*10MeV}{\Delta V}$ at which the heat just reaches to the
border of the pearl. That is to say for smaller parameter values
there is a significant loss in energy to higher frequencies, while
for larger values of our parameter we expect that a major part of the
energy from the contraction manages to be emitted as the line.

We obtain a good fit to the frequency and intensity of the 3.5 keV line
for a pearl of fixed mass M = 1.4*$10^8$ GeV
with the parameter value $\frac{\xi*10MeV}{\Delta V} = 4.2$. The
corresponding radius and surface tension of the pearl are:
\begin{equation}
R = 2.8\, cm \qquad S^{1/3} = 6.7\, GeV.
\end{equation}


\section{Latest
 Idea: Smaller Pearls givng also DAMA
Observation and
Tycho Supernova Remnant Observation of 3.5 keV}
\label{small}
After the Bled conference we have looked at the idea that we could
ignore the connection to the Tunguska event, which was at first so
terribly important for our studies and instead seek a combined
fitting of not only as just presented the 3.5 keV radiation from the
clusters of galaxies and the center of our Milky Way, but also
an observation, that would at first look like spoiling the
hypothesis that the 3.55 keV line comes from dark matter.
In fact this observation  was considered by the authors of \cite{Jeltema} to be
a clear sign that the 3.5 keV  line must after all be an effect of some
ordinary ions - such as an unexpectedly high abundance of potassium (K) -
but not a signal from dark matter. This observation is the observation
by Jeltema and Profumo that the 3.5 keV line is indeed also emitted from
the Tycho supernova remnant! In almost all usual dark matter models as
elementary particles this appearance from the
supernova remnant with very little dark matter compared to ordinary matter
is rather absurd. It can only come about if the dark matter can
somehow absorb the energy present in the remnant region and convert it into
the 3.5 keV line.

We are now working on fitting the requirement to get sufficient
3.5 keV radiation from the supernova remnant and it certainly points towards
smaller values for the tension than even the fit above.

In fact we have a crude fit to both the observation by Jeltema and Profumo
and the DAMA and DAMA LIBRA observations, but now with both the
cubic root of the tension $S^{1/3}$ and the potential difference for a
nucleon passing through the skin of the pearl $\Delta V$ being of the order
of 1 or 2 MeV only.

In this picture the pearls are less than atomic size and thus
much more like
dark matter models
with WIMPs. But, especially to cope with the amount of interaction needed
for the Tycho supernova remnant observation, they have to interact
so strongly that they will interact several times on the way down
through the earth to the DAMA-LIBRA observatory. So they should not be called
weakly interacting, i.e the W in WIMP should be left out. Because they are,
however, still very heavy, say $10^3$ GeV or even heavier, compared to usual
WIMP speculations, they are difficult to stop even when they hit
matter in the shielding. So they can pass on and penetrate into the
apparatus even if they have been somewhat hitting matter on the way down.
Assuming that they as macroscopic objects - they are
 still pearls
although now smaller - have somewhat different cross sections,
some pearls may come through.
Then even if only a small part comes through the shielding they could
cause a number of events,  as the observations suggest anyway
in experiments like DAMA-LIBRA.
Actually such a survival is only expected for some exceptional ones
among the dark matter particles, which could easily
lead to an enhanced dependence on the season and thus be especially
suitable to be detected by DAMA-LIBRA relative to
other experiments, that just observe the events independent of
their seasonal variation.

\subsection{$\frac{\sigma}{M}$ from Tycho Observation}
\label{Tycho}
The mysterious 3.55 keV line has been seen, corrected to zero Doppler shift,
not only from various
galaxy clusters and the Milky Way Center, but also
from the remnant of the supernova described by Tycho Brahe after its appearance
in 1572. This at first seems to be in contradiction
to the hypothesis that the X-rays should come from dark matter at all.

The authors Jeltema and Profumo \cite{Jeltema} take it that this Tycho supernova
remnant observation means that the 3.55 keV line radiation  cannot come from
dark matter because basically there would not be dark matter in sufficient
amounts in the supernova remnant. It would then have to be an ordinary transition
line in excited ions, which must have been underestimated in the theoretical
calculation of the other radiation from the supernova remnant say.
Actually some underestimate of the abundance of potassium K could deliver
a line in the region.

But we basically take the point of view, that dark matter consists of
some (type of) particles which have the possibility of being excited, and then
when excited to send out especially X-rays in the 3.55 keV line. So we have the
option of having the activity in the supernova remnant excite the dark matter
particles there and thus make them radiate with their characteristic
frequency 3.55 keV. (In the galactic clusters etc. we have a model of exciting
them by collisions causing skin contraction and thus extra energy being
set free. But the emission is again the characteristic line 3.55 keV.)

But of course the absolute imperative for such a model for creating the 3.55 keV line
radiation in the supernova remnant is that the dark matter particles (whatever
they may be) have sufficiently big cross sections to at least pick up enough energy
for the emission of the observed 3.55 keV line radiation.

\subsubsection{How we got the need for
$\frac{\sigma}{M}\ge  6*10^{-7}m^2/kg$}

\subsubsection{What observed:}
\label{or}
Jeltema and Profumo claim \cite{Jeltema} that they have observed an X-ray
spectral peak - fitted with difficulty, but nonetheless fitted to be there -
with an intensity of $2.2*10^{-5}$ photons per $cm^2$ per $s$.
Thus in each $cm^2$ of the sphere around Tycho passing
through the earth, there passes $2.2 *10^{-5}$ photons per s per $cm^2$.

The distance to Tycho (SN1572) is about 9000 light-years.
In fact, according to Wikipedia:

``The distance to the supernova remnant has been estimated to between
2 and 5 kpc (approx. 6,500 and 16,300 light-years), with recent studies
suggesting a narrower range of 2.5 and 3 kpc
(approx. 8,000 and 9,800 light-years).''

Taking 1 light-year $= 10^{16}$ m,
the  area of the  sphere around Tycho going through the earth
is
\begin{eqnarray}
sphere \: area &=& 4\pi *( 9000 ly * 10^{16} m/ly)^2\\
&=& 10^{41} m^2
\end{eqnarray}

So the number of 3.55 keV photons passing through this surface will be
\begin{eqnarray}
\# \: of \: photons &=& (2.2\pm 0.3)*10^{-5}cm^{-2}s^{-1}*10^{41}*10^4cm^2\\
&=& 2 * 10^{40}s^{-1}\\
&\sim& ``an \:  energy \: rate'': 3.5 keV *2*10^{40}s^{-1}\\
&=& 10^{32} erg/s.
\end{eqnarray}

\subsubsection{Rate of Energy Ploughing up}

The total energy in the remnant region will still in first
approximation be equal to the energy ejected from the
supernova, if we assume that the energy escaping as light going
so far away that we no more can count it as belonging to
the remnant is small compared to the part remaining in the
remnant region. A major part of this energy is presumably in the form
of fast moving particles or even X-rays, so that order of magnitudewise
we may count it as cosmic rays moving with the speed of light
relative to the dark matter pearls, which of course have a much
lower velocity of the order of the escape velocity from the Galaxy.

All over the remnant region we assume that the density of dark matter
is very similar to that in the neighborhood of our solar system
\begin{eqnarray}
D_{sun} &=& \frac{0.3\, GeV}{cm^3},
\end{eqnarray}
so that the number of pearls we have in every $cm^3$ is
$\frac{0.3\, GeV}{M}$. In each second each of these pearls pick up
the cosmic rays or whatever
material in the remnant in a volume $\sigma *v \approx \sigma*c$
where $\sigma$ is the cross section for a pearl and $v$ is the average
relative velocity of the pearl and the remnant matter or radiation.
That is to say, that during a second the fraction of the volume
getting ploughed through is
\begin{eqnarray}
\hbox{``Fraction ploughed through''} &=& \frac{D_{sun}}{M}*\sigma*v.
\end{eqnarray}
So if one observes a 3.55 keV line with an intensity
$I = 2.2 *10^{-5}$ photons per s per $cm^2$
we need the total energy rate (power) at a distance
$d = 9*10^{19}m$ to be
\begin{eqnarray}
W &=& I *4\pi * d^2*3.55\, keV*
2.2*10^{-5}\, cm^{-2}s^{-1}
= 10^{32}\, \frac{erg}{s}.
\end{eqnarray}
Then, if all the energy is converted into 3.5 keV radiation, we must have
\begin{eqnarray}
W&=&E_{remnant}* vD_{sun}*\frac{\sigma}{M},
\end{eqnarray}
and the lower bound for $\frac{\sigma}{M}$ is
\begin{eqnarray}
\left. \frac{\sigma}{M}\right |_{Tycho} &=& \frac{W}{E_{remnant}*D_{sun}v}
\label{eq28}\\
&=& \frac{10^{32}erg/s}{10^{51} erg *0.3\, GeV/cm^3 *3*10^{10}cm/s}\\
&=&0.56 *10^{-2}cm^2/kg\\
&=& 10^{-29}cm^2/GeV\\
&=& \frac{1}{(3.4\, GeV)^3}
\label{eq32}
\label{bound}
\end{eqnarray}

\subsection{Comparing to Nuclear $\sigma/M$ Ratio}
\label{coincidence}
The material inside our pearls is highly
compressed and taken to be mainly carbon (with atomic number $A =12$). Then using a
crude formula $1.2 A^{1/3} fm$ for the radius of a nucleus and
$\pi (1.2 A^{1/3})^2 fm^2$ for the cross section for some smaller particle
scattering on the nucleus, we get for nucleus scattering:
\begin{eqnarray}\label{nuclear}
\left .\frac{\sigma}{M}\right |_{nuclear}&=& \frac{\pi * 1.2^2 fm^2* A^{2/3}}
{A*0.94\, GeV} \\
&=&  \frac{123\, GeV^{-3}}{\sqrt[3]{12}}\\
&=& \frac{1}{(0.265\, GeV)^3}.
\end{eqnarray}
Combining these numbers for the ratio $\frac{\sigma}{M}$ needed
for the dark matter in the supernova remnant (\ref{bound}) with the one
for a suitable nucleus (\ref{nuclear}) we see that the
needed lower bound is
\begin{eqnarray}
\frac{\left .
\frac{\sigma}{M}\right |_{Tycho}}{\left. \frac{\sigma}{M}\right |_{nuclear}}
&=& \frac{(0.26\, GeV)^3}{(3.4\, GeV)^3}\\
&=& 0.076^3 = 4.5 *10^{-4}.
\end{eqnarray}
This means that about 1/2000 of the
accessible energy would indeed become 3.5 keV
photons, if the cross section for the pearls in Tycho was actually equal to the
nuclear cross section. Actually such an efficiency of $4.5 *10^{-4}$
is not at all unlikely. So we could claim that, having in mind that
the orders of magnitude could have run out to wildly different
values, the rather close agreement could be taken to mean that indeed the
true $\frac{\sigma}{M}$ for the dark matter pearls being excited
is indeed equal to the nuclear one (\ref{nuclear}). If indeed the pearls
were so small that there was no significant shadowing by one nucleus of
another of the nuclei in the pearls, then the cross section to mass ratio
would just be the nuclear one. So an order of magnitude agreement with
the actual cross section to mass ratio being the nuclear value should be
taken almost as successful agreement.


Our original lower bound (\ref{eq28}) to (\ref{eq32}) is calculated
under the assumption that the speed of the particles bringing the energy
is equal to the speed of light and that all the available energy goes into
the 3.55 keV line. However we shall group the ingredients of the supernova
remnant into two parts, neither of which contribute 100\% of their energy
to the line and only one part, the cosmic radiation, has approximately
the speed of light.

Let us consider the two components of the remnant-material separately:
\begin{itemize}
\item{Cosmic Rays}

In order to explain the cosmic rays observed coming
{\bf from} the supernova remnant, we need of the order of
5\% to 19\% of the energy in the remnant to be contained in the  cosmic rays.

If we think of a pearl of $10^4$ GeV mass and thus with of the order
of $10^4$ electrons it takes of the order $10^4*3.55\, \hbox{keV} =40$ MeV
energy to  heat the pearl to the temperature 3.55 keV. So for a cosmic ray
with more than 40 MeV energy hitting the pearl the temperature will, after some
time $t_{spread}$, rise above the 3.55 keV even at the surface and then
the emission of the line frequency 3.55 keV will be beaten out
in the competition with the higher frequencies. This causes a suppression
of 3.55 keV line emission
by a factor $\frac{t_{spread}}{t_{radiation}}$, to be compared with
what is supposed to happen in the dark-matter + dark matter collisions
in which the emitted energy $E_S$ is of the order of a few per mille of the Einstein mass,
which we just took to be $10^4$ GeV, meaning say $E_S \sim 30$ GeV. So only the
lowest tail of the cosmic ray energy spectrum down at and below
40 MeV is useful for producing large amounts of 3.55 keV radiation.

If we surmise from looking at the cosmic ray spectrum,
which is rather flat for low energy up to about 1 GeV where then the
famous decrease by a third power sets in, that the fraction not getting
strongly suppressed by the time ratio factor is about 1/20, then the useful
cosmic ray energy is about 1\% of the total energy.

But this part, this 1 percent, now risks being lost to
lower then 3.55 keV frequencies. In fact the excitons, which are supposed by their
decay to deliver the 3.55 keV radiation, will competitively decay into
phonons instead of the photon line.
We argue below in subsection \ref{promise} that
the photon emission going to the line 3.55 keV is suppressed relative
to the phonon-type of decay of the exciton by a
factor $\alpha$, the fine structure constant.

So crudely we estimate that from the cosmic ray part of the supernova remnant
only about $\frac{1\%}{137}\approx 1/14000$ of the energy goes to the 3.55 keV line.

This alone corresponds to delivering  1/7 of the radiation, which would be observed
if the $\frac{\sigma}{M}$ was what we called the ``nuclear ''
value. So within our crude estimating this would in itself
be enough to fit the hypothesis that the true cross section to mass
ratio {\em is } the ``nuclear'' one.

\item{Gas or Plasma}

But these particles rather have
velocities about 100 times slower than the speed of light \cite{Williams}.

The gas and the plasma in the remnant has a typical velocity
of the order 3000 km/s = 1\% of the speed of light \cite{Williams}.
That of course already reduces
the contribution from the gas and the plasma by a factor 100 compared
to particles with light speed.
Each of these collisions only bring
about $(1\%)^2$ GeV = 100 keV to the pearl, which is below the above mentioned
40 MeV. Thus there is essentially no energy loss to high frequency radiation,
but only to radiation with lower frequency than the 3.55 keV line.

In fact when we have this low total energy situation and the temperature
even locally falls below 3.55 keV,
excitons decay into phonons as well as to the 3.55 keV line.
As we shall go into below there is then a factor $\alpha$ suppressing
the 3.55 keV line decay mode relative to the phonon decay mode.

So, if there was no further suppression, the conversion rate of
energy into 3.55 keV radiation for the gas and plasma would
be suppressed by the combined factor
$1\% *1/137 \sim 1/14000$ compared to the maximally allowed rate. This is
accidentally the same factor we got for the cosmic rays.

But actually there is a further suppression in as far as
for the charged ions there is an electric field around the pearl
pushing positively charged particles away. The relevant energy barrier
is $\Delta V \sim 2$ MeV, which is more than the 100 keV expected for the
gas or plasma impacts. So only the neutral atoms may come
through the skin of the pearl.
However even the deflected protons, say, can
excite the electron system of the pearl by the
Coulomb interaction and the energy lost by a
proton in a collision can be used for producing radiation of the 3.55 keV line.
Anyway, if the number of neutrals
is similar in order of magnitude to the number of ions, then this last suppression
may at the end not be so significant.

\end{itemize}

The 3.55 keV radiation from both components of the material in the supernova remnant
seems to be suppressed relative to the ideal full use of the energy
by about the same factor 1/14000. This means that the true cross section over mass
ratio has to be about 14000 times bigger than the limit we first calculated.
We found that the ``nuclear'' cross section to mass ratio
is about 2000 times bigger than the bound 
in the paper by Jeltema and Profumo \cite{Jeltema}.

It really seems that a reasonable estimate can support the idea that the true
cross section to mass ratio could be what we called the nuclear one.

\begin{figure}
\includegraphics{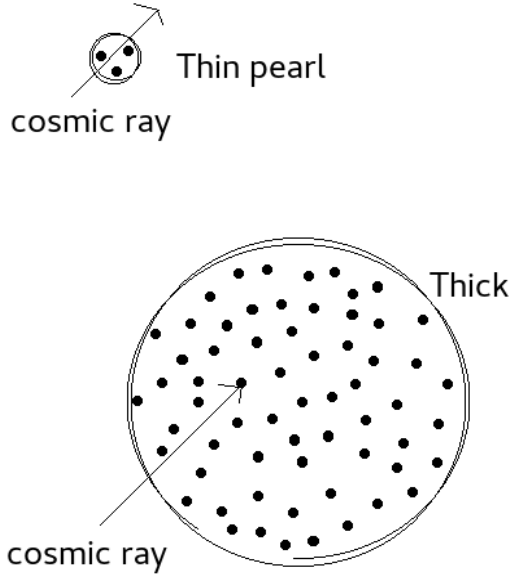}
\caption{This figure illustrates that for the density inside a pearl
being very high a cosmic ray particle hitting the pearl will
sooner or later in the interior hit a nucleus, while for a very little
pearl with the same density the thickness of the pearl is insufficient
for all cosmic ray particles to hit a nucleus and the cross section
will be less than the geometrical one $\sigma= \pi R^2$. The ratio
$\frac{\sigma}{M}$ is then rather equal to the nuclear value (\ref{nuclear}).}
\label{thickness}
\end{figure}

The special likelihood that the true ratio should be
the ``nuclear one'' is explained here again in other words:


If we assume that the tension $S$ and the parameter
$\frac{\xi_{fS}}{\Delta V}$ have such values that
{\em formally} the cross section to mass ratio $\frac{\sigma}{M}$
would be smaller than the corresponding nuclear ratio (\ref{nuclear}),
the actual cross section to mass ratio would only be
approximately equal to the nuclear ratio. (Here $\xi_{fS}$ is the
radius scaling factor for fixed tension $S$, see section \ref{xifs}).
However for sufficiently thin pearls a cosmic ray say could with high probability pass
through the pearl without hitting any nuclei inside.
For such parameters one would obtain for the
cross section to mass ratio just the nuclear value, see Figure \ref{thickness}.


Optimistically we could say that our estimate of the efficiency of
the conversion of energy in the supernova remnant into 3.55 keV radiation
suggests that the
size of the pearls is actually so small that the cross section to mass ratio
becomes equal to the nuclear ratio. But for this to happen it would have to
be that the formally calculated ratio should be {\em larger} than or
equal to this nuclear ratio. This in turn will
put an upper limit on the tension $S$ depending somewhat on our parameter
$\frac{\xi_{fS}}{\Delta V}$, since the cross section to mass ratio
is a decreasing function of the tension $S$ and then of course also as
a function of the third root of this tension $S^{1/3}$ which we mainly
use in our text and figures.

\subsubsection{The promised $\alpha$-factor.}
\label{promise}

For the above mentioned factors $\alpha$ it is crucial that the decay rate into
the 3.55 keV line is of the order $\alpha$ lower than the rate of an exciton
decaying into phonons.

Of course if we consider the radiative decay with a 3.55 keV photon being
emitted, there must be an $\alpha$ factor in the decay
rate.

However we argue that in the low temperature situation considered the
displacements of the lattice in the phonon-modes
have sizes which do not depend in a simple way on the fine structure constant,
but are rather just given by the elasticity properties of the material,
the sound velocity and the density. We take the electron single particle
energy eigenstates to be close to momentum eigenstates and not
of a form depending severely on $\alpha$.
The lattice displacements will then modify the wave functions and
give some overlap between the hole and the electron wave functions
from which the exciton decay rate into phonons may be estimated. The point
now is that such decay rates do not contain an explicit factor of alpha.

Note that in the above argument we have to keep in mind that the homolumo gap
is very small compared to the fermi momentum or energy say. Therefore
in the decay of an electron-hole pair with such a small energy - as the homolumo gap -
only a small range of momentum shifts of the electron
is possible. This range may though be bigger for the phonon decay mode
than for the X-ray emission because of the lower velocity
of the phonon than the photon. Ignoring this difference in range
for the two modes under the small energy in the decay, it will indeed
turn out that the photon emission is lower in rate by a factor
of order $\alpha$.


\subsection{Combined Fitting, Small Pearl Model}
\label{CF}
\subsubsection{Formulas for the Critical Case, Pearls Just about to Collapse}

First let us give a list of the interesting quantities in terms of the
cubic root of the tension of the surface $S^{1/3}$ and the energy difference
for the nucleon on passing the domain wall $\Delta V$ in the case
of a critical sized pearl. By this we mean the case in which
a further parameter has been avoided by adjusting it so that the tension
provides a pressure on the material inside the pearl making it just on the
border to collapse by spitting out nucleons. In other words providing enough pressure
to just barely compensate the potential difference $\Delta V$ per nucleon.
So now we should note the various parameters in this borderline/critical
situation (see reference \cite{theline} for details on the notation):
\begin{eqnarray}
\hbox{Pearl radius } R_{crit} &=&
\frac{3 \pi^2S}{2 (\Delta V)^4} \\
\hbox{Fermi momentum } p_{f \; crit} &=& 2\Delta V\\
\hbox{Energy release by collision } E_{S \; crit} &=&  S(\sim 4\pi) R_{crit}^2\\
&=&\pi^5*9S^3/(\Delta V)^8\\
\hbox{Collision cross section } \sigma_{crit} &=&\pi*(2R_{crit})^2=
9*\pi^5
S^2/(\Delta V)^8  \\
t_{spread \; crit}&=& \frac{\rho c_p}{4k}*R^2|_{crit}\\
&=& \frac{\alpha 55 R^2 T}{24 c^3}|_{crit}\\
t_{radiation\; crit } &=&
\frac{E_S}{4\pi R^2\sigma_{ST} (3.5 keV)^4}\\
&=& \frac{60 S}{\pi^2 (3.55 keV)^4};\\
\frac{\sigma_{crit}}{M_{crit}}&=& \frac{
9\pi^5
S^2/(\Delta V)^8}{ m_N *\frac{24 \pi^5 S^3}{(\Delta V)^9}}\\
&=& \frac{\Delta V   }{
\frac{8}{3}S m_N}\\
\frac{E_{S \; crit}}{M_{crit}}
&=& \frac{S(\sim 4 \pi) (\frac{3 \pi^2 S}{(\Delta V)^4})^2 }
{m_N \frac{24 \pi^5 S^3}{(\Delta V)^9}}\\
&=& \frac{\Delta V }{\frac{8}{3}m_N}\\
\frac{N_{crit}}{M_{crit}}= \frac{E_{S \; crit}}{M_{crit} *3.55\, keV}&
=&
\frac{\Delta V}{
\frac{8}{3}m_N *3.55\, keV}\\
\left.\frac{N\sigma}{M^2}\right |_{all E_S \rightarrow 3.5 keV; \; crit}&=&
\frac{N_{crit}}{M_{crit}}*\frac{\sigma_{crit}}{M_{crit}}\\
&=& \frac{(\Delta V)^2}{
\frac{16}{3}S m_N^2*3.55\, keV}\\
 \frac{t_{spread}}{t_{radiation}}
*\left.\frac{N\sigma}{M^2}\right |_{all E_S \rightarrow 3.5 keV; \, crit}
&=& \frac{1.23*10^{-15}\, MeV}{\Delta V^5}\\
frequency=E_H &=&137^{-3/2}\sqrt{2}p_f =
137^{-3/2}\sqrt{2}2\Delta V\\
\sqrt[3]{\frac{9\pi M_{crit}}{8 m_N}}&= & (Rp_f)|_{crit} \\
\frac{M_{crit}}{m_N} &=& \frac{24 \pi^5 S^3}{(\Delta V)^9}.
\end{eqnarray}

\subsubsection{With Radius Scale up Parameter $\xi_{fS}$}
\label{xifs}
The critical case is not realistic except very crudely. The pearls
would collapse by the tiniest deformation during the contraction in the early
universe situation. We must expect that there must be an appreciable
safety margin in the sense, that the number of nucleons inside the contracting
pearl for the pearl not to collapse immediately must be so large, that the
final radius, when the fluctuations from the contraction have died out will be
say $R = \xi_{fS}*R_{crit}$ with $\xi_{fS} \approx 5$. We estimated in earlier
articles \cite{theline} this expected ratio of the average radius to the critical or borderline
one to be $\sqrt{4\pi}*2^{4/9} \approx 5$.

The dependence of some of the important quantities with this $\xi_{fS}$
goes as follows. Here we also include the dependence on $\Delta V$ and on $S$:

\begin{eqnarray}
\hbox{Pearl radius } R &=& \xi_{fS}R_{crit}= \xi_{fS}\frac{S*24 \pi^2}
{(2\Delta V)^4} \\
\hbox{Cubic root of tension } S^{\frac{1}{3}} &=& S^{1/3}\hbox{(fixed)} \\
\hbox{Fermi momentum } p_f &=&  \xi_{fS}^{-1/4}2\Delta V\\
\hbox{Energy release by collision } E_S &=&\pi^5*9S^3\xi_{fS}^2/(\Delta V)^8\\
\hbox{Collision cross section } \sigma &=&
9\pi^5\xi_{fS}^2S^2/(\Delta V)^8  \\
t_{spread}
&=& \frac{\alpha 55 R^2 T}{24 c^3}\\
\hbox{(where $T\approx 0.3 \Delta V$)}\\
&=& 1.10 \Delta V *S^2  \left ( \frac{\xi_{fs}^{1/4}}{\Delta V}\right)^8\\
t_{radiation }
&=& \frac{60 S}{\pi^2 (3.55\, keV)^4}\\
&=& \frac{6.08 S}{(3.55\, keV)^4};\\
\frac{t_{spread}}{t_{radiation}}
&=& \frac{ 1.10 \Delta V *S^2  \left ( \frac{\xi_{fs}^{1/4}}{\Delta V}\right)^8}
{\frac{6.08 S}{(3.55\, keV)^4}}\\
&=& 0.18*(3.55\, keV)^4*S\left (\frac{\xi_{fS}^{1/4}}{\Delta V}\right )^8
\Delta V\\
\frac{\sigma}{M}&=& \frac{
9*\pi^5
S^2\left (\frac{\xi_{fS}^{1/4}}{\Delta V}\right )^8}
{ m_N *24 \pi^5 S^3\left ( \frac{\xi_{fS}^{1/4}}{\Delta V} \right )^9}\\
&=& \frac{1}{
\frac{8}{3}S m_N*\frac{\xi_{fS}^{1/4}}{\Delta V}} \label{71}\\
\frac{E_{S}}{M}&=& \frac{S(\sim 4\pi) R_{crit}^2}
{ m_N24 \pi^5 S^3\left (\frac{\xi_{fS}^{1/4}}{\Delta V}\right )^9}\\
&=&  \frac{1}{\frac{8}{3}m_N*\frac{\xi_{fS}^{1/4}}{\Delta V} }\\
\frac{N}{M}= \frac{E_{S}}{M *3.55\, keV}&
=&
\frac{1}{
\frac{8}{3}m_N *\frac{\xi_{fS}^{1/4}}{\Delta V}*3.55\, keV}\\
\left .\frac{N\sigma}{M^2}\right |_{all E_S \rightarrow 3.5\, keV}&=&
\frac{N}{M}*\frac{\sigma}{M}\\
&=& \frac{1}{
\frac{16}{3}*\frac{4}{3}S m_N^2*\left ( \frac{\xi_{fS}^{1/4}}{\Delta V}\right )^2
3.55\, keV}\\
 \frac{t_{spread}}{t_{radiation}}
*\left.\frac{N\sigma}{M^2}\right |_{all E_S \rightarrow 3.5 keV}
&=&
0.18*(3.55\, keV)^4S\left (\frac{\xi_{fS}^{1/4}}{\Delta V}\right )^8
\Delta V * \nonumber\\
&&* \frac{1}{
\frac{16}{3}*\frac{4}{3}S m_N^2*\left ( \frac{\xi_{fS}^{1/4}}{\Delta V}\right )^2
3.55\, keV} \label{intensity}\\
&=&
1.23*10^{-15}\, MeV \Delta V *\left ( \frac{\xi_{fS}^{1/4}}{\Delta V}\right )^6\\
frequency=E_H &=&137^{-3/2}\sqrt{2}p_f =
\xi_{fS}^{-1/4}\frac{2\sqrt{2}}{137^{3/2}}\Delta V \label{frequency}\\
\frac{M}{m_N}&= & \frac{8}{9\pi}(Rp_f)^3\\
&=&
24 \pi^5\left ( \frac{S^{1/3}\xi_{fS}^{1/4}}{\Delta V}\right )^9 \label{82}
\end{eqnarray}

For the fitting of the intensity we use (\ref{intensity-exp}) and (\ref{intensity}):
\begin{eqnarray}
8.1*10^{-4}\, GeV^{-4}&=&
1.23*10^{-15}\, MeV\Delta V*
\left ( \frac{\xi_{fS}^{1/4}}{\Delta V}\right )^6\\
\hbox{or with $\Delta V =10\, MeV$ say}&&\\
\left ( \frac{\xi_{fS}^{1/4}}{\Delta V}\right )^6&=& \frac{8.1*10^{-16}\, MeV^{-4}}
{
1.23*10^{-14}\, MeV^2}=
0.0659\, MeV^{-6}\\
\hbox{ giving } \frac{\xi_{fS}^{1/4}}{\Delta V}&=& \sqrt[6]{
0.0659\, MeV^{-6}}\\
&=& 0.635\, MeV^{-1}
\end{eqnarray}

Similarly for the frequency fitting of $E_H = 3.55\, keV$ we use (\ref{frequency}):
\begin{eqnarray}
\frac{\xi_{fS}^{1/4}}{\Delta V}&=& (3.55\, keV)^{-1}*\frac{2\sqrt{2}}{137^{3/2}}\\
&=& 0.50\, MeV^{-1}.
\end{eqnarray}

Thus we fit the frequency and intensity of the X-ray line with essentially
the same value of our parameter $\frac{\xi_{fS}^{1/4}}{\Delta V}$. So our
small pearl model provides a good fit to the astronomical data.

Using $\left. \frac{\sigma}{M}\right |_{Tycho} = \frac{1}{(3.4 \; GeV)^3}$
and equation (\ref{71}) we have:
\begin{eqnarray}
\frac{1}{(3.4 \; GeV)^3} &=&  \frac{1}{
\frac{8}{3}S m_N*\frac{\xi_{fS}^{1/4}}{\Delta V}}
\end{eqnarray}
or
\begin{eqnarray}
(S^{1/3})^3*\frac{\xi_{fS}^{1/4}}{\Delta V} &=& (3400\, MeV)^3*3/(8m_N)\\
&=& 1.57*10^7\, MeV^2
\end{eqnarray}
E.g. we can ask: what is $S^{1/3}$, if we want $\frac{\xi_{fS}^{1/4}}{\Delta V}
=0.50\, MeV^{-1}$? The answer is $S^{1/3}= \sqrt[3]{3.14*10^7\, MeV^3}$
= 315\, MeV.

If we instead assume the ``nuclear'' value for $\frac{\sigma}{M}$, which is
2000 times bigger we get of course from (\ref{71}) a $\sqrt[3]{2000}$ times
smaller $S^{1/3} = 25$ MeV. This last value is, however, only to be considered
an upper bound for $S^{1/3}$;
if the tension
$S^{1/3}$ was even smaller the $\frac{\sigma}{M}$ ratio would not
increase further, because the say cosmic ray particles hitting the
pearl would penetrate the pearl and effectively only experience
the ``nuclear'' value. Thus we really have
\begin{eqnarray}
S^{1/3} &\le & 25\, MeV.
\end{eqnarray}
With the fitted value $\frac{\xi_{fS}^{1/4}}{\Delta V} = 0.5\,
MeV^{-1}$
this corresponds according to equation (\ref{82}) to an upper limit for the
mass
\begin{eqnarray}
M & \le & 5.1 *10^{13}\, GeV
\end{eqnarray}
and an upper limit for the radius
\begin{eqnarray}
R &\le & 2.9 *10^{-9}\, m
\end{eqnarray}
ensuring in itself the upper limit from the DAMA-LIBRA
experiment $M \le 1.56 *10^{14}$ GeV.


\subsection{DAMA-LIBRA Mass Extraction}
\label{DAMA}
The major speculation and idea behind the small pearl study, in addition
to the inclusion of the Jeltema and Profumo observation of 3.55 keV
X-ray radiation from the Tycho supernova remnant, is the inclusion of an
attempt to fit and explain the controversial DAMA-LIBRA experiment\cite{DAMA}.
In contrast to other underground searches for dark matter, DAMA-LIBRA
{\em did} find the dark matter by the technique of seasonal variation.

According to the discussion in subsection \ref{coincidence}
the cross section to mass
ratio $\frac{\sigma}{M}$ for our pearls needed to fit reasonably the
Tycho supernova remnant observation agrees - we wanted to say as
a ``coincidence'' (but that is only very optimistically true) - with
the same ratio for e.g. carbon nuclei.

Indeed we found (\ref{nuclear})
\begin{eqnarray}
\left . \frac{\sigma}{M}\right |_{nuclear} &=&\frac{1}{(0.26\, GeV)^3}\\
&=& 1.25 *10^{-3}m^2/kg \label{nuclearvalue}
\end{eqnarray}
while the DAMA-LIBRA experiment
presented two allowed regions for WIMP observation
in the mass of the particle versus cross section plane:
\begin{eqnarray}
(M, \sigma) &=& ( 18\, GeV, 2*10^{-4}\, pb)= (3.2*10^{-26}\,kg , 2*10^{-44}\, m^2)\label{DAMAmass1}
\end{eqnarray}
and
\begin{eqnarray}
(M, \sigma) &=& ( 180\, GeV, 10^{-4}\, pb)= (3.2*10^{-26}\, kg , 10^{-44}\, m^2),\label{DAMAmass2}
\end{eqnarray}
giving respectively
\begin{eqnarray}
\frac{\sigma}{M} &= & \frac{2*10^{-4}\, pb}{18\, GeV}\\
&=& 6.24 *10^{-19}\, m^2/kg
\end{eqnarray}
and
\begin{eqnarray}
\frac{\sigma}{M}&=& \frac{10^{-4}\, pb}{180\, GeV}\\
&=& 3.1 *10^{-20}\, m^2/kg.
\end{eqnarray}

 It means that the ratio $\frac{\sigma}{M} $ fitted to WIMPs by DAMA
 is about a factor $10^{12}$ (or even $10^{13}$) lower than the number
which our fit using the Jeltema and Profumo 3.55 keV observation
points to, namely $6*10^{-7}m^2/kg$.
If we take it that really the $\frac{\sigma}{M}$ ratio
for our pearls is equal to the nuclear value, then the deviation from
the observed ratio in DAMA-LIBRA is even larger, by about a factor
2000 bigger.

As we shall see in the next subsubsection \ref{M1} we estimate that the number
of particles / events observed
requires that the mass be at most $1.56 *10^{14}$ GeV,
since otherwise with the known density $D_{sol} \approx 0.3\,GeV/cm^3$ there
could not be enough particles so as to fit the observed ones.


The main idea now is that our dark matter pearls have a rather high cross section
and thus cannot avoid interacting with the large amount of earth above the DAMA
detector. Due to the filtering and braking of the pearls we assume that this shielding
of the detector effectively removes all but one particle in $10^{12}*\frac{M}{M_{fit \, DAMA}}$.
This number was just taken from the comparison of the assumed cross section and
the seemingly measured one giving a ratio $10^{12}$ as we just discussed and
then correcting it by the factor $\frac{M}{M_{fit \: DAMA}}$.
Here $M_{fit \; DAMA}$ stands for the mass of the WIMP fitted by the
DAMA-LIBRA group. To cope with this suppression of the
number of particles we need an increase in the number coming in by the
factor $10^{12}*\frac{M_{fit \; DAMA}}{M}$. Thus we have to use this factor
to reduce our estimate of the pearl mass M relative to the
$1.56*10^{14}$ GeV to obtain:
\begin{eqnarray}
M= \hbox{``Mass estimate from DAMA''} &\approx& \frac{1.56*10^{14}\, GeV}
{10^{12}*\frac{M_{fit \; DAMA}}{M}}\\
&=& 160\, GeV*\frac{M}{M_{fit \; DAMA}}
\end{eqnarray}
or
\begin{eqnarray}
M_{fit \; DAMA} &=& 160\, GeV.\label{droppedout}
\end{eqnarray}
This mass of 160 GeV is in very good agreement with the
above values (\ref{DAMAmass1}, \ref{DAMAmass2}) of 18 GeV or 180 Gev.
So it is very well
consistent with the DAMA-LIBRA data that the $\frac{\sigma}{M}$
ratio should be equal to the lower bound needed for the Tycho supernova
remnant observation by Jeltema and Profumo.

Let us, however immediately mention a little correction:

The fit by the DAMA-LIBRA experimentalists of course assumed that the
modulated part of the signal they observed was only a small part of
all the dark matter hits they saw. However, in our rather IMP-model
it is most likely that the modulated signal is a much larger part of the
full signal. Crudely we would argue that the modulated signal in an
IMP model is likely to be of the same order of magnitude as the full signal.
In other words the main part of the unmodulated signal would in fact be background.
If now the fit of the DAMA-LIBRA had taken such a point of view
the formal cross section they would deliver would have been
appreciably smaller by a factor equal to the ratio between the
bulk and the modulated signal. With this correction it then
means that the suppression would be stronger by such a ratio and the
factor $10^{12}$ which we used would be even bigger by something like
two orders of magnitude. That in turn would make the $\frac{\sigma}{M}|_{true}$
ratio, having been measured so to speak by DAMA-LIBRA, would go up from
the  one agreeing with the lower bound from Tycho to a bigger value
much closer to the nuclear one, although not quite there.


Had we used the supposed more correct value by taking the $10^{12}$
bigger by a factor 2000, or rather including the mass
correction $10^{12}*2000*\frac{M_{fit \; DAMA}}{M}$,
we would only get consistency provided we have the proposed extra correction
by claiming that in truth the modulation effect is almost 100\% of the dark
matter signal.

But note that since the mass $M$ dropped out of the equation (\ref{droppedout})
we derived, this consideration left the mass $M$
of the pearls unrestricted. However we had of course already found
an upper limit for the mass $M$, namely
\begin{eqnarray}
M &\le & 1.56*10^{14} GeV.
\end{eqnarray}

Furthermore
our pearls must not just be ordinary atoms surrounded by a skin, they
must be {\em many} atoms surrounded by the skin. There shall at least be so
many, $Z$, charges on protons in the pearl that a potential of the order of
magnitude of $\Delta V$ can be achieved. Using the well-known formula for the
ground state of the electron binding energy for a hydrogen like atom
$ZRy$ $\approx$  $Z *13$ eV, we need to get $Z \approx 10^5$ at least, just
to reach even the surprisingly small $\Delta V \sim 1$ MeV coming out of our fit to the
small pearls. So we cannot keep the model unless we let the mass $M$ of the
pearls be at least $10^5$ GeV.

Combining these bounds for the mass, it must be in the interval:
\begin{eqnarray}
1.56 *10^{14}\, GeV &\ge \quad M \quad\ge & 10^5\, GeV.\label{ineq}
\end{eqnarray}


\subsubsection{How Many Particle Hit the DAMA Experiment?}
\label{M1}
In this subsubsection we shall now estimate the promised
 approximate absolutely lowest needed number of dark matter
particles coming in and thereby the upper bound on the mass of these
particles as follows:

The modulated part of the signal is found by DAMA/LIBRA
to be of the order 0.01 cpd/kg/keV in the region of
energy of the signal in the range 1keV to 6 keV where any modulation
if found at all. Taking this as averaged over the range of 5 keV
it means that one in total saw at least 0.05 cpd/kg even modulated
and thus dark matter related events meaning for the whole apparatus
about $250\, kg*0.05\, cpd/kg = 12.5\, cpd$. Since the apparatus has an area
of the order of $ 1/4 \;  m^2$ - it consists of 25 essentially
$10\times 10\times$...blocks - this means an absolutely needed flux
- whatever the theory - of $50\, cpd/m^2$. Here cpd means counts per day,
and should be compared to what we trust about the dark matter:
We have in our region a mass density $ 0.3\, GeV/cm^3= 3*10^5\, GeV/m^3$ and a
velocity of the order 300 km/s meaning 300 km/s *86400s/day
= $26*10^6$ km/day = $2.6* 10^{10}$ m/day. So 1/4 $m^2$ tracks
per day a volume $1/4 *2.6 *10^{10} m^3$ =$6.5 *10^9m^3$ containing a mass of
$6.5 *10^9 m^3 * 3*10^5\, GeV/m^3 = 19.5 *10^{14}\, GeV = 2.0 *10^{15}\, GeV$.
This $ 2.0 *10^{15}\, GeV$ mass is to be shared on 12.5 counts,
since there have been seen 12.5 cpd. Thus the particles must at
least have masses less than or equal to $2.0*10^{15}\, GeV/12.5
=1.56 *10^{14}\, GeV$.
There is the possibility that with the strongly interacting
pearls in our small mass model the modulation part relative to the total
number of interactions with the apparatus gets appreciably enhanced.
In fact the depth into which the pearls penetrate must be strongly
dependent on the impact velocity, since it takes more collisions
to stop a fast pearl than a slow one (compared to Earth velocity).
Since presumably the DAMA-LIBRA experiment is working with the
few pearls coming especially deep down the number of them could be
very strongly velocity dependent.
It is in fact possible that these modulation part particles
are almost the only dark matter particles, although this would usually be
a bit strange if it were so.
Such enhancement of the modulation could explain the long standing
mystery, why DAMA-LIBRA sees the dark matter while the
other experiments - not using the modulation technique - do not
see anything.




\subsection{Simple Formulas on Underground Searches for Dark
Matter}
\label{simple}
Usually people assume that dark matter consists of weakly interacting
particles, so called WIMPs (= weakly interacting massive particles). But
if the particles could be heavy, they could also
be so strongly interacting that the particles would
interact several times on the way down through the earth
shielding the experiments looking for dark matter
underground. However they do not need to be sufficiently strongly
interacting that it would make them visible on the sky.
Such particles would not deserve the
name WIMP but rather only IMP.

Since all we know from the gravitational effect of the dark matter is
the mass density $D$, the quantity that crudely measures the
degree of visibility of the dark matter would be the amount of
absorption or of any kind of observable effect, say some cross section
$\sigma$ per unit volume in outer space. For fixed $D$ that quantity
would be proportional to the ratio $\frac{\sigma}{M}$, i.e.
to the amount of cross section per unit mass.

We shall in this section, taking just this ratio $\frac{\sigma}{M}$,
look for what one crudely measures in experiments looking for
WIMPs or IMPs impacting on earth.

Calling the mass of the average nucleus or whatever is taken to be
the most important constituent of the earth hitting the dark matter
particles $M_{nucleus}$, we may crudely estimate that the number of
collisions it takes for a dark matter particle to be effectively
stopped in passing through the shielding is
\begin{eqnarray}\label{Nhits}
\hbox{`` Number hit for stop''}&\approx& \frac{M}{M_{nucleus}}.
\end{eqnarray}
 The argument for this estimate is the following:

During its passage through the shielding - the layer of earth above the detector -
the dark matter particle / pearl  of mass $M$ hits earth particles of mass
$M_{nucleus}$, which then obtain a speed of the order of magnitude of the speed
$v$ of the dark matter particle itself. Thereby the hit particles achieve
a kinetic energy of the order of $M_{nucleus}v^2/2$ which is
$\frac{M_{nucleus}}{M}$ times the kinetic energy of the dark matter pearl itself
$Mv^2/2$. Thus to bring the kinetic energy of this pearl down to about zero
it is needed of the order of the inverse of the fraction $\frac{M_{nucleus}}{M}$
such hits. But that is just what (\ref{Nhits}) says.

\subsubsection{Estimation of Number of Hits Needed}

As we shall see in a moment we shall avoid the pearl making too many hits
when passing the counting sensitive region of the experiment. The reasons are:
\begin{itemize}
\item If one sees more than one hit in the experiment, one counts it
as a background interaction and does not include it in the usual
searches for WIMPs.
\item Below we shall give an estimate of the number of hits to be seen in
the experimental sensitive region. If there are many interactions/hits in
this region there will not be so many counts of something happening as the
estimation below. They will so to speak be used up on multiple hits
instead.
\end{itemize}

We estimate now an effective thickness of the experimentally sensitive region
in say the DAMA-LIBRA experiment to be of the order of $l_{sensitive}=\frac{1}{2}\, m$.
Then we argue that the stopping length $l_{stop}$ divided by
`` Number hit for stop'' $\approx \frac{M}{M_{nucleus}}$,
should be larger than or of order of magnitude of $\frac{1}{2}\, m$. I.e.
\begin{eqnarray}
\frac{l_{stop}M_{nucleus}}{M} &\ge& l_{sensitive}\approx \frac{1}{2}\, m.
\end{eqnarray}

\subsubsection{Penetration in Terms of $\frac{\sigma}{M}$}

If one thinks of WIMPs the very number of observed dark matter
particles or pearls in an underground experiment is proportional
(crudely at least) to the ratio $\frac{\sigma}{M}$ of cross section
to mass. This is because, taking the density of dark matter $D$ in the astronomical
neighborhood and the typical velocity $v$ as given,
the flux of dark matter particles passing by becomes inversely
proportional to the mass $M$ and the interaction rate must of course always
be proportional to the cross section $\sigma$ for hitting.

Therefore really the ratio $\frac{\sigma}{M}$ estimated by an underground experiment
is basically an estimate of the intensity of hits in the sensitive part of the
apparatus. Assuming dark matter consists of WIMPs this number is basically measured by the
underground experiments, essentially just by counting events.

Now, however, if the pearls interact several times on their way down through
the shielding then the effect of such full or partial stopping of the
particles can of course drastically change the result of measuring the ratio
$\frac{\sigma}{M}$ as if they were WIMPs.

Almost by dimensional arguments we could write down the stopping length
\begin{eqnarray}
l_{stop} &=& \frac{M}{\sigma \rho_{shield}}.
\end{eqnarray}
In fact supposing that the shielding material is mass-wise dominated by the
one particle - presumably a nucleus - of mass $M_{nucleus}$ the (mass) density
is given as
\begin{eqnarray}
\rho_{shield}&=& \hbox{``number density''}*M_{nucleus}.
\end{eqnarray}
and the distribution of the pearl's first hit on this material is given as
\begin{eqnarray}
& \propto & \exp(- l_{hit}x) \hbox{ (where $x$ is depth into shielding)}
\end{eqnarray}
where
 \begin{eqnarray}
  l_{hit}&=& \frac{1}{\hbox{``number density''}*\sigma}\\
&=& \frac{M_{nucleus}}{\rho_{shield}*\sigma},
\end{eqnarray}
 we obtain
\begin{eqnarray}\label{lstop}
l_{stop}&=& \frac{M}{M_{nucleus}}*l_{hit} \label{lstoplhit}\\
&=& \frac{M}{\sigma \rho_{shield}}.
\end{eqnarray}

For simplicity we shall at first assume that the suppression
of the rate of the part of the dark matter coming through
the shielding is proportional to $\exp(-x/l_{stop})$ where $x$ is the depth,
meaning the penetration depth into the earth, even in
the case of multiple scattering.
This simplification is of course not mathematically true
and we shall return to it later.
However proceeding with our simplifying assumption
we find that the
cross section to mass ratio $\left. \frac{\sigma}{M}\right |_{WIMP}$ to be
effectively found as if we had WIMPs
will be


\begin{eqnarray}
\left. \frac{\sigma}{M}\right |_{as \; WIMP}*\frac{M_{as \; WIMP}}{M}
&=& \frac{\sigma}{M}*\exp(-x/l_{stop}).\\
\hbox{ or } \sigma|_{as \; WIMP} &=& \sigma_{true} *\exp(-x/l_{stop})
\end{eqnarray}
Using  (\ref{lstop}) we write this in the form
\begin{eqnarray}\label{transcendent}
\left. \frac{\sigma}{M}\right |_{as \; WIMP}*\frac{M_{as \; WIMP}}{M}&=&
\frac{\sigma}{M}*\exp(-\frac{x\rho_{shield}\sigma}{M}),
\end{eqnarray}
which we can consider as a transcendental equation from which to
determine the true $\frac{\sigma}{M}$ for the dark matter
pearls from the experimentally observed ``as if WIMP''
value $ \left. \frac{\sigma}{M}\right |_{as \; WIMP}$, which can be
identified with the DAMA-LIBRA fitted value. There is in this equation
for a small value of the $ \left. \frac{\sigma}{M}\right |_{as \; WIMP}$
the WIMP-solution, but there are {\em two} solutions. The second solution
is a strong coupling solution. To solve the equation in this strong coupling
case we of course have to put in the value of the depth $x$ under earth of the
experiment. It is given as 3400 mwe (= meters water equivalent),
which means we can put $x=3400\, m$ and then $\rho_{shield}=1000\, kg/m^3$.
In principle we have to correct for the fact that the dark matter particles
will typically move in a skew direction and the true value of $x$ will be somewhat
larger than the minimal distance from the earth's surface to the experiment.
Since we anyway
calculate very crudely and since in the strongly interacting case the
shortest way down will come to give the dominant contribution,
we here simply take $x=3400\, m$ and $\rho_{shield} =1000\, kg/m^3$.
Then we obtain
\begin{eqnarray}
(x\rho_{shield})_{for \; DAMA} &=& 3400\, m *1000\, kg/m^3\\
&=& 3.4*10^6 \, kg/m^2
\end{eqnarray}
For illustration let us remark that e.g. for what we called
``nuclear'' cross section to mass ratio
$1.25 *10^{-3}m^2/kg $, see equation (\ref{nuclearvalue}), the exponent
would become
$-3.4*10^6kg/m^2 *1.25 *10^{-3}m^2/kg = - 4.3 *10^{3}$.

 The cross section to mass ratio for
WIMPs seemingly observed in the
DAMA-LIBRA controversial underground experiment may be taken from the
 two allowed regions
in the mass of particle versus cross section plane
as presented by the experimentalists:
\begin{eqnarray}
(M, \sigma) &=& ( 18\, GeV, 2*10^{-4}pb)= (3.2*10^{-26}kg , 2*10^{-44} m^2)
\end{eqnarray}
and
\begin{eqnarray}
(M, \sigma) &=& ( 180\, GeV, 10^{-4}pb)= (3.2*10^{-26}kg , 10^{-44} m^2),\label{M180}
\end{eqnarray}
giving respectively
\begin{eqnarray}
\left .\frac{\sigma}{M}\right |_{as \; WIMP} &= &
\frac{2*10^{-4}pb}{18\, GeV}=
\frac{2*10^{-44}m^2}{3.2*10^{-26}kg}\\
&=& 6.24 *10^{-19}m^2/kg
\end{eqnarray}
and
\begin{eqnarray}
\left .\frac{\sigma}{M}\right |_{as \; WIMP}&=& \frac{10^{-4}pb}{180\, GeV}=
\frac{10^{-44}m^2}{3.2 *10^{-25}kg}\\
&=& 3.1 *10^{-20}m^2/kg
\end{eqnarray}

Solving the transcendental equation (\ref{transcendent}) iteratively
we first find that
\begin{eqnarray}
x\rho_{shield} *\frac{\sigma}{M} &\approx & \ln \left ( \frac{\sigma}{M}
* \left . \frac{M}{\sigma} \right|_{as \; WIMP} *\frac{M}{M_{as \; WIMP}} \right ).
\end{eqnarray}
Taking at first the logarithm to be of order unity we shall test as first
iteration
$\frac{\sigma}{M} = (x\rho_{shield})^{-1} = (3.4 *10^6kg/m^2)^{-1}$ =
$2.94 *10^{-7}m^2/kg$. But inserting that value into the logarithm
gives the value $\ln (\frac{2.94*10^{-7}m^2/kg}{10^{-20}m^2/kg}/
\frac{M_{as \; WIMP}}{M} )$
= $\ln (3*10^{13}/ 0.1) =33 $.
Here we have used $M_{as\, WIMP}$ = 180 GeV (\ref{M180}) and M = 2000 GeV (\ref{M2000})
giving $\frac{M_{as \; WIMP}}{M} \sim $ 0.1.

So the next iteration gives
\begin{eqnarray}
\left. \frac{\sigma}{M}\right |_{2. \; sol.}&\approx & 2.94*10^{-7}m^2/kg *33\\
&=& 1.0*10^{-5} m^2/kg
\end{eqnarray}

Crudely we can consider this number
$1.0*10^{-5} m^2/kg $ as the DAMA
measured value for the cross section to mass ratio provided the second
- i.e. the strong interaction solution - is taken.

This value is then to be compared to the value we need for the
Jeltema and Profumo Tycho supernova observation:
\begin{eqnarray}
\frac{\sigma}{M}|_{Tycho}&=& 5.6 * 10^{ -7} m^2 /kg. 
\end{eqnarray}
The ``measured'' value is only
18 times larger than the
one required for the Tycho supernova remnant observation.

But remember now we speculated that this number from
the Tycho observation is only a lower limit and that we
suggested the $\frac{\sigma}{M}$ ratio should be a factor
2000 times bigger than the Tycho measurement.
Such a factor as that would bring the deviation from the ``measured ratio''
to the opposite side. So we should really conclude that
the DAMA estimation of the ratio and that from Tycho are in agreement.

\subsubsection{Number of Hits during Stopping}

The number
$33$ which we got for the value of the logarithm in the
solving of the transcendental equation above is actually equal to the depth
$x$ measured in stopping lengths $l_{stop}$. And so we would conclude
that
\begin{eqnarray}
33*\rho_{shield}l_{stop} &=& (3400\, mwe)* \rho_{water} = 3.4 *10^6kg/m^2,
\end{eqnarray}
giving
\begin{eqnarray}
\rho_{shield}l_{stop} &=&  \frac{3.4 *10^6 kg/m^2}{
33} =
1.0*10^5 kg/m^2.
\end{eqnarray}
Now in order
to avoid getting more than one
hit in the sensitive thickness of the apparatus taken to be 1/2 m,
we have the inequality:
\begin{eqnarray}
l_{hit}&\ge & \frac{1}{2}\, m.
\end{eqnarray}
So taking the density in this sensitive apparatus to be say
$ \rho_{apparatus}=3000\, kg/m^3$, we have
\begin{eqnarray}
\frac{\rho_{apparatus}l_{hit}}{\rho_{shield}l_{stop}}&\ge &
\frac{(
33*3000\, kg/m^3)* \frac{1}{2} m }{3400\, m *1000\, kg/m^3}\\
&=&
1.5*10^{-2} = \frac{1}{
69}.
\end{eqnarray}
Using $\rho_{apparatus} \approx \rho_{shield}$ and equation (\ref{lstoplhit})
this means
\begin{equation}
\frac{M_{nucleus}}{M} \ge \frac{1}{69}.
\end{equation}
If say the important or average nucleus in the shield is silicon with
mass 28 GeV, then the pearl's mass is of the order of
\begin{equation}
M  = 69*28\, GeV = 2000\, GeV.\label{M2000}
\end{equation}

\subsubsection{Requirement for Proper Macroscopic Physics}

\begin{figure}
\includegraphics{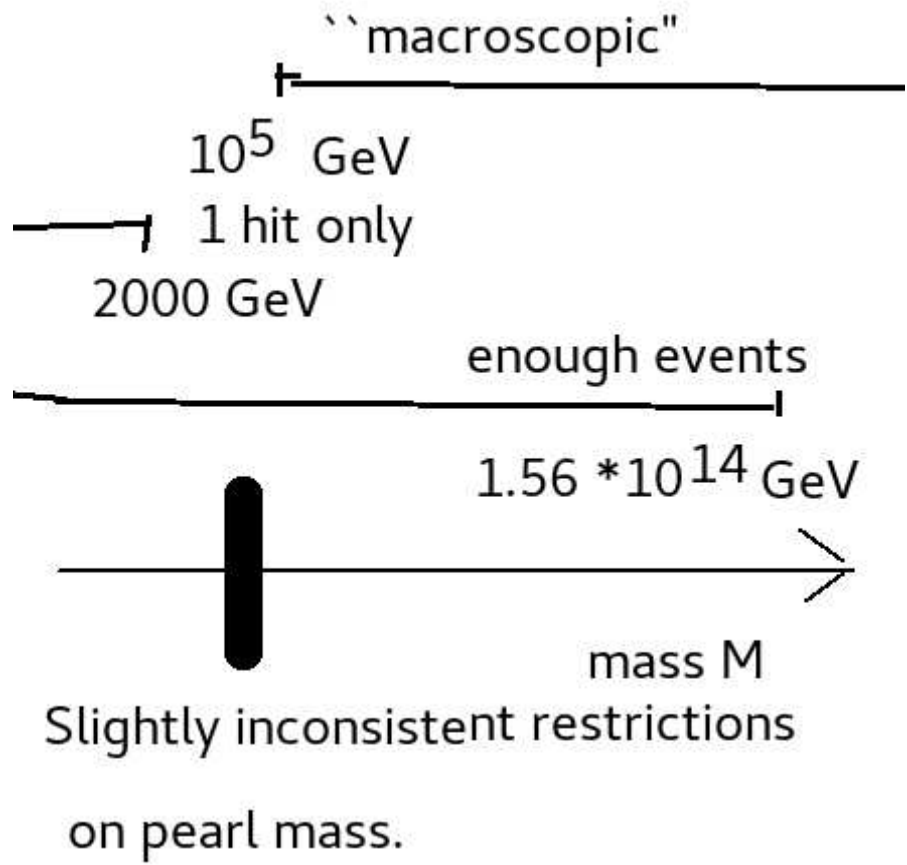}
\caption{Three slightly inconsistent requirements for the mass $M$
of the pearl: The requirement $M\ge 10^5$ GeV comes from the condition
that there should be enough electrons present so that you can treat the
bulk of the pearl as a macroscopic piece of $\sim $ metal;
the requirement $M\le 2000$ GeV comes from the condition
that no more than one hit in the DAMA sensitive region is seen at a time;
the upper bound $M\le 1.56 *10^{14}$ GeV comes from requiring enough events
as counted by DAMA-LIBRA. The three bounds are slightly inconsistent, but
within uncertainties a value $M\approx 10^4$ GeV should be considered a
good fit.}
\label{mass2}
\end{figure}

Now in order to have a proper macroscopic electron cloud in the pearl
that can give the macroscopically estimated homolumo gap, we need that
the pearl nuclear charge $Z$  (i.e. the number of protons) is at least
large enough that
an atom of this atomic number can provide $\Delta V$ order of magnitude
binding energies. Taking the binding energy to be of the order
of $Z$  Rydberg, it means we need $Z \ge \frac{\Delta V}{ 1 Rydberg}$,
so that for say $\Delta V = 1$ MeV we would need $Z\ge 10^5$.
This would be a problem for our model if we took the above  estimate
of 2000 GeV too accurately.
But this limit is so close that we shall of course rather take it
that now we know the bound must be very close and we shall take
the mass to lie in the range $M \approx 2000$ to $100,000$ GeV - see Figure
\ref{mass2}. As a reasonable compromise we shall take
$M \approx 10^4$ GeV.

\subsubsection{An Interesting Coincidence}
Let us note, that we have got almost coincidence between
the mass $10^5$ GeV needed for our macroscopic approximation
to be valid and the value obtained above (\ref{M2000}).
In other words we can say that the mass needed for
keeping a sufficiently high electron density such
that e.g. our homolumo-gap calculation is still valid and the
mass estimated from DAMA-LIBRA, say 2000 GeV,
are essentially the same, which is a funny coincidence!

Actually if we begin to fit with a mass a bit smaller than $10^5$ GeV,
there will be a correction to the formula for the
homolumo gap size and thus for our prediction of the very
frequency 3.55 keV. So the true prediction of this
frequency would be a bit lower, if such corrections for the bigger extension
of the electron cloud than the size of the skin is corrected for.

This actually means that the true homolumo gap has a maximum very near to the
values we here use to fit with. This may be of some significance for
really getting a peak in the X-ray spectrum (at 3.55 keV), since
a priori pearls of a bit different size will give different
frequencies for the radiation and thus smear out the peak relative to
what would appear, if all the pearls have exactly the same size.
It may only go with the fourth root of the radius scaling factor
$\xi_{fS}$ that there is such a dependence
but still it is a smearing out.

Suppose it happens that the dominant size of the pearls is just around
a point where the approximation of the electron cloud keeping inside
the skin of the pearls stops being valid. Then there will be a correction that
for making the pearl smaller counteracts the increase in frequency that the
smaller pearl should cause. The result is a maximum in the frequency spectrum of the
X-ray radiation. This means an improvement in the
sharpness of the line is predicted.

If we somehow argue that just such a maximum is favoured it would
mean we could consider this coincidence as a success.

\subsection{Xenon1T Electron Recoil Excess}
\label{xenon1t}
An observation that may fit very well into our version of the
pearl model for dark matter with the less than atomic size pearls
is the Xenon1T Electron Recoil Excess \cite{Xenon1Texcess}. This effect of electrons seemingly
appearing with energy close to just 3.5 keV - note the coincidence we want to
stress with the 3.5 keV X-ray line photon energy - would independent of the details
of the dark matter model be very indicative, since we already have a strong
suggestion that dark matter tends to emit light with the 3.5 keV
frequency.

Apart from the DAMA/LIBRA and DAMA experiment the other direct search experiments
seem to find only negative results when looking for the dark matter.
However one unexpected result \cite{Xenon1Texcess} was found, although
at first not seemingly due to dark matter:

The experiment Xenon1T investigated what they call electron recoil.
In the Xenon experiment one has a big tank
of liquid Xenon with some gaseous Xenon above it and photomultipliers
looking for the scintillation of this xenon. The philosophy behind the experiment is
that a dark matter WIMP e.g. hits a nucleus inside the xenon
and the recoil of this nucleus creates a scintillation signal S1 and also an electron
which is then driven up the xenon tank by an electric field.
At the end the electron is made to give a signal S2 in the gaseous
Xenon at the top by a further electric field.
By the relative size of the signals S1 and S2 one may classify the events
- which are taken to be almost coinciding pairs of these signals S1 and S2 -
as being nucleus recoil or electron recoil. One expects to find
the dark matter in the nucleus recoils, since a dark matter particle
is not expected to make an electron with sufficient energy to make an
observable electron recoil event.

But now, by carefully estimating the expected background, the Xenon1T
experimenters found an excess of electron recoil events.
Proposed ideas for explaining it include axions from the sun or neutrinos having
bigger magnetic moments or perhaps less interestingly that there could be more
tritium than expected in the xenon.

But here our model of relatively stronger interacting particles
able to radiate the line 3.55 keV when excited provides a possible explanation:

Going through the earth and the rest of the shielding the pearls or
particles get excited so as to emit a 3.55 keV X-ray just as they
would do in the Tycho supernova remnant, where they also get
excited by matter or cosmic rays. But then the particles passing through the
deep underground Xenon1T experiment are already excited and
prepared for sending out the 3.55 keV radiation. Now they could possibly
simply do that in the xenon tank or they might dispose of the energy
by a sort of Auger effect by rather sending out an electron with an
extra energy of 3.55 keV. Such an electron with an energy of a few keV
could be detected and taken for an electron recoil event in the
Xenon1T experiment.

It is remarkable that the signal of these excess electron recoil events
appears to have just an energy of the recoiling electron very close to the
value 3.55 keV. Indeed the most important bins for the
excess are the bins between 2 and 3 keV and the bin between 3 and 4 keV.

So we would claim that in our model there is no need for extra
solar axions, a bigger neutrino magnetic moment or tritium.
But we claim it to be 3.55 keV radiating dark matter that one sees in the xenon
experiment!

\section{Conclusion}
\label{conclusion}
We have put up two slightly different models for dark matter being actually
pearls which have a new phase or type of vacuum inside. By our
``Multiple Point Principle'' this new vacuum is supposed to have the same energy
density as the present vacuum. The two models only differ by taking the parameters
different, especially the tension of the surface separating the inside with
its vacuum from the outside with the present vacuum.

The two models are thus given as roughly:
\begin{itemize}
\item Big pearls, adjusted to the Tunguska event being due to one falling
down onto the earth:

The cubic root $S^{1/3}$ of the tension is several GeV, the size of the
pearls is cm-size.

\item Small pearls:

The cubic root of the tension $S^{1/3}$ is of the order of
1 MeV, the size of the pearls is a bit bigger than atomic nuclei.
\end{itemize}

Our main result was that we could fit both the very frequency 3.5 keV
of the X-ray radiation suspected to come from dark matter and the
intensity as fitted by Cline and Frey to a series of observations of this
line from various galaxy clusters with essentially one parameter.
We wrote this parameter as $\frac{\xi *10MeV}{\Delta V}$ for large
pearls or $\frac{\xi_{fS}^{1/4}}{\Delta V}$ for small pearls. So two observed quantities
are fitted by one parameter. Both observations concern the still doubtful
3.5 keV X-ray radiation.

We can essentially fit with this parameter whether we take the pearls
big with a big surface tension or small with a small surface tension.

Taking the model with the small pearls, on which we have far from finished
everything, we hope that we can further:
\begin{itemize}
\item Make the DAMA-LIBRA
controversial observation of dark matter by the seasonal
variation technique compatible with the model.
\item Fit the a priori very strange observation by Jeltema and Profumo
of 3.5 keV radiation coming from the Tycho supernova remnant
in the picture with the 3.5 keV radiation
coming from dark matter.
(Something they take themselves as the sign that this 3.5 keV line is
not coming from dark matter but from some ion such as potassium).
\item We have for our model a very promising coincidence of the
electron excess energy from the Xenon1T experiment with the number 3.5 keV.
The point is that our pearls - in the small size model -
come through the apparatus of the Xenon1T experiment and are excited
with some extra electrons or simply have some excitons in them
- excited during the passage through the shielding - which then deliver just
 the 3.5 keV energy to an electron in the Xenon1T experiment. And that is
then giving an excess of such events with just an excited electron which was
the unexpected effect seen by Xenon1T.
\end{itemize}

\subsection{The fitting of the Small Pearl Version}
We basically make predictions from the
small pearl version with the following parameters:
\begin{itemize}
\item The surface tension represented by its cubic root: $S^{1/3}$,
\item Essentially the potential difference $\Delta V$ for a nucleon
inside versus outside the pearl, represented by the combination
$\frac{\xi_{fS}^{1/4}}{\Delta V}$. Here $\xi_{fS}$ is  the ratio of the
radius of
the pearl to the ``critical'' radius at which the nucleons would be just
about to be spit out. Presumably even coming in under the fourth root
this ratio $\xi_{fS}$ is not of much significance and probably is $\sim 5$.
\end{itemize}

We found that the mass of the pearl should lie
inside the rather large interval (\ref{ineq})
\begin{eqnarray}
1.56 *10^{14}\, GeV &\ge \quad M \quad\ge & 10^5\, GeV, \label{massrange}
\end{eqnarray}
except that a slightly weak argument using the DAMA-LIBRA experiment and
assuming our pearls not to scatter more than once led to a mass
$M \approx 2000 GeV$, barely consistent with the allowed interval.
The mass range (\ref{massrange}) corresponds to taking
the cubic root of the tension $S^{1/3}$ parameter in the range
\begin{eqnarray}
28\, MeV  &\ge \quad S^{1/3} \quad  \ge \quad 2.7\, MeV.
\end{eqnarray}
So in some sense we only used the $\Delta V$ or rather
the $\frac{\xi_{fS}^{1/4}}{\Delta V}$ parameter, meaning one parameter.

But then we combined our small pearl model with the assumption that the
parameter $S^{1/3}$ was
small enough that the pearls became
so small
that they are sufficiently transparent for say nuclei that the effective ratio
of cross section to mass $\frac{\sigma}{M}$  for hitting nuclei becomes
equal to that for say carbon nuclei, a ratio we called the ``nuclear''
cross section to mass ratio. This hypothesis turned out to fit remarkably
well with
the Tycho supernova remnant observation! It also fits
very well the value of this ratio extracted from the DAMA-LIBRA experiment.



Further it is a coincidence, although not obviously reasonable to understand
physically, that the size of the pearls is just such that the electron cloud
begins to emerge significantly outside the skin surrounding the pearl.
This means
that the homolumo gap providing the very frequency 3.55 keV for the
radiation has a maximum at just this fitted situation. Thus
the 3.55 keV line will be especially sharp compared to the possibility
that this coincidence was not realized.


\subsection{Parameters $S^{1/3}$ and $\Delta V$ Small and Outlook}

The parameter values we obtained with our ``Small Pearls Version''
for a pearl mass $M \sim 10^4$ GeV are
\begin{eqnarray}
S^{1/3} &=& 2\, MeV\\
\frac{\xi_{fS}^{1/4}}{\Delta V} &=& 0.5\, MeV^{-1},
\end{eqnarray}
which with
\begin{eqnarray}
\xi_{fS}&\approx & 2^{4/9}*\sqrt{4\pi}\approx 5
\end{eqnarray}
gives
\begin{eqnarray}
 \Delta V &\approx & 1.3\, MeV.
\end{eqnarray}
The corresponding radius of the pearl is
\begin{equation}
R = 1.5*10^{-12} \, m.
\end{equation}

The parameters $S^{1/3}$ and $\Delta V$
are - one would say embarrassingly -
small compared to the dimensional argument expectations, if one
speculated that Higgs physics and top-quark physics were involved.
That would namely instead give e.g. $S^{1/3} \sim 100$ GeV. This means
that Higgs and/or top-quark physics is not at all a promising
possible explanation behind the vacuum-phases. We rather need
physics of an energy order of magnitude even under or at least in the
very low energy scale end of strong interaction physics, or it
should be rather a kind of atomic physics involved.

We have ideas under development taking as a starting point
the work by Kryjevski  Kaplan and Schaefer \cite{KKS}, who
calculated the phase diagram for nuclear matter under various
high nuclear densities and considered the
so called CFL phase. This stands for color flavour locking phase
meaning that the $SU(3)_c$ color group is broken spontaneously in a direction
locked with that of the flavour $SU(3)_f$ group. It is remarkable that these
authors find a triple point as a function of the light quark masses coinciding
with the experimental quark masses. This is, however, not quite what we would
need to have a case of MPP degenerate {\em vacuum}-phases. Because of the
high baryon density used in the study of Kryjevski  Kaplan and
Schaefer \cite{KKS} their phases are namely not vacua.

Nevertheless we are working on the idea that their phase diagram might be
extrapolated down to zero baryon density and thus tell us about
vacuum phases. In that case an energy scale for the phase transition physics
of the order of the strong interaction scale $\Lambda_{QCD}
\approx 300 MeV$ could be understandable. Even
reaching down to a few MeV is at least closer than if one should begin with the
Higgs-mass scale.

Such surprisingly low tension domain walls also bring the chances for them
to really be acceptable astronomically much closer. The problem with
domain walls coming to dominate energetically the whole cosmology and thus
being phenomenologically unacceptable is of course weakened the lower the
tension and thereby from Lorentz invariance also the energy per unit wall-area
is.

\end{document}